\documentclass[aps,prd,superscriptaddress,showpacs,showkeys,floatfix,preprintnumbers,nofootinbib,letterpaper]{revtex4-2}
\usepackage{setspace}
\usepackage{verbatim}
\usepackage{tikz}
\usetikzlibrary {arrows.meta}
\usepackage{color}
\usepackage{amssymb}
\usepackage{pgfplots}
\usepackage{amsmath,amssymb}
\usepackage{mathtools}
\usepackage{hyperref}

\usepackage{graphicx}  
\usepackage{dcolumn}   
\usepackage{bm}        
\usepackage{amssymb}   
\usepackage{physics}
\usepackage{mathtools}
\usepackage{ulem}

\definecolor{Rocco}{rgb}{0.0, 0.5, 0.0}
\definecolor{Resolved?}{rgb}{0.5, 0.0, 0.5}
\newcommand{\mc}{\mathcal}

\newcommand{\QQbar}{{Q\bar Q}}
\newcommand{\Qbar}{{\bar Q}}

\newcommand{\mcD}{{\mathcal D}}
\newcommand{\FTEE}{FTE$^2$ }

\newcommand{\lp}{\left}
\newcommand{\rp}{\right}

\newcommand{\mcA}{{\mathcal A}}
\newcommand{\mcAbar}{{\overline\mcA}}
\newcommand{\mcE}{{\mathcal E}}
\newcommand{\mcG}{{\mathcal G}}
\newcommand{\mcH}{{\mathcal H}}
\newcommand{\mcM}{{\mathcal M}}
\newcommand{\mcO}{{\mathcal O}}

\newcommand{\mcZ}{{\mathcal Z}}

\begin{document}
\title{Entanglement entropy of a color flux tube in (2+1)D Yang-Mills theory}
\author{Rocco Amorosso}
\affiliation{Department of Physics and Astronomy, Stony Brook University, Stony Brook, NY 11794, USA}
\author{Sergey Syritsyn}
\affiliation{Department of Physics and Astronomy, Stony Brook University, Stony Brook, NY 11794, USA}
\author{Raju Venugopalan}
\affiliation{Physics Department, Brookhaven National Laboratory, Upton, NY 11973, USA}
\affiliation{CFNS, Department of Physics and Astronomy, Stony Brook University, Stony Brook, NY 11794, USA}

\begin{abstract}
We construct a novel flux tube entanglement entropy (FTE$^2$), defined as the excess entanglement entropy relative to the vacuum of a region of color flux stretching between a heavy quark-anti-quark pair in pure gauge Yang-Mills theory. We show that FTE$^2$ can be expressed in terms of correlators of Polyakov loops, is manifestly gauge-invariant, and therefore free of the ambiguities in computations of the entanglement entropy in gauge theories related to the choice of the center algebra. Employing the replica trick, we compute FTE$^2$ for $SU(2)$ Yang-Mills theory in (2+1)D and demonstrate that it is finite in the continuum limit. We explore the properties of FTE$^2$ for a half-slab geometry, which allows us to vary the width and location of the slab, and the extent to which the slab cross-cuts the color flux tube. Following the intuition provided by computations of FTE$^2$ in (1+1)D, and in a thin string model, we examine the extent to which our FTE$^2$ results can be interpreted as the sum of an internal color entropy and a vibrational entropy corresponding to the transverse excitations of the string.    

\end{abstract}

\date{\today}

\maketitle
\tableofcontents

\clearpage

\section{Introduction}

In quantum field theories, fields in a given spacetime region are strongly entangled with fields outside it. 
The entanglement entropy of a region $V$ in a quantum system 
is determined from the reduced density matrix
\begin{equation}
\label{eqn:rho_reduced}
\hat{\rho}_V=\Tr_{\bar V}\hat{\rho}\,,
\end{equation}
where $\hat{\rho}$ is the full density matrix of the system and $\Tr_{\bar V}$ represents the partial trace over quantum fields in the region $\bar V$ outside $V$.
The (von Neumann) entanglement entropy is then 
\begin{equation}
\label{eqn:VNEE}
   S_E=-\Tr \lp(\hat{\rho}_{V} \log \hat{\rho}_V\rp)\,.
\end{equation}
The entanglement entropy thus defined  diverges logarithmically in $D=2$ dimensions with the smallest resolvable ultraviolet scale of the theory and is proportional to the area of the region for $D > 2$ ~\cite{Bombelli:1986rw,Srednicki:1993im}.
This scaling of the entanglement entropy has been argued to have a geometric origin corresponding to the 
``deficit angle" of the region of interest relative to the excised region outside~\cite{Callan:1994py}; in the context of black holes, it can be understood as a quantum correction to the classical Bekenstein-Hawking entropy.
Powerful renormalization group approaches (such as the C-theorem, originally proposed for 2D conformal field theories~\cite{Zamolodchikov:1986gt}) provide insight into the nature of the microstates that constitute the entanglement entropy.
A comprehensive review of entanglement entropy in numerous contexts can be found in \cite{Nishioka:2018khk}.

Given the spacetime structure of entanglement entropy, it is natural to ask whether aspects of the dynamics of confinement in gauge theories can be understood in the language of quantum information.
In 2D QED and Yang-Mills theories, confinement is clearly observed to have a stringy structure~\cite{Coleman:1975pw,tHooft:1974pnl} and its relation to entanglement entropy and quantum information has been widely explored, with much of the recent interest driven by the promise of quantum computing.
For gauge theories in higher dimensions, the relation of confinement to quantum information is far more subtle. 
This is in part because, due to the dynamical nature of the gauge field, constructing entanglement measures while respecting gauge invariance is not straightforward. We will discuss this point at length in this paper. The other reason the relation of confinement to entanglement is nontrivial is because it is not clear one can assign in general geometrical structure to confinement~\cite{Jokela:2020wgs}. 

Considering the latter aspect first, it widely believed that large $N_c$ Yang-Mills theory in $D > 2$ dimensions has also a stringy structure~\cite{tHooft:1973alw} allowing one to interpret confinement similarly to the $D=2$ case~\cite{Polyakov:1980ca}. A further significant development was the development of holographic techniques (via the AdS/CFT conjecture) to directly compute the entanglement entropy for a class of nontrivial gauge theories~\cite{Ryu:2006bv,Ryu:2006ef}. The connection of this holographic entropy to the confinement-deconfinement transition was noted in \cite{Witten:1998zw} and developed further in \cite{Klebanov:2007ws} to explore its geometric structure at both zero and finite temperature. 

With regard to pure glue Yang-Mills theory for $D> 2$, and in QCD, the best tool to explore this connection is lattice gauge theory. There is ample numerical evidence from the lattice\footnote{Experiments can also be interpreted as providing support for stringy underpinnings of phenomena, despite string breaking in QCD. One example is the nearly linear behavior seen in hadron Regge trajectories~\cite{Bali:2000gf}. Further, a semi-classical LUND string model first developed to describe the spacetime structure of meson production in $e^+e^-$ collisions~\cite{Andersson:1983ia}, captures many features of multi-particle production at high energy colliders~\cite{Sjostrand:2019zhc}. Interestingly, quantum entanglement in the Schwinger model description of the expanding quark-antiquark string provides an explanation for the apparently puzzling thermal character of the particle spectrum~\cite{Berges:2017zws,Berges:2018cny}.} for the existence of stringy structures~\cite{Bali:2000gf}. In particular, the potential between a pair of heavy quarks in the fundamental representation can be described by a confining chromoelectric color flux tube with a string tension $\sqrt{\sigma}\sim 440$ MeV \cite{Teper:1997am}. 
In particular, the energy levels between the heavy quarks can be described by a relativistic string and its transverse excitations~\cite{Luscher:1980iy,Luscher:1980fr,Arvis:1983fp}. For detailed lattice studies of string dynamics in $SU(N_c)$ Yang-Mills, we refer readers to \cite{Bali:1994de,Athenodorou:2007du,Cardoso:2013lla,Athenodorou:2021qvs}; many of its features are reproduced in an effective field theory (EFT) of long strings derived from the Nambu-Goto action~\cite{Polchinski:1991ax,Aharony:2009gg,Dubovsky:2012sh,Aharony:2013ipa}. 

The entanglement entropy for  $SU(2)$ Yang-Mills theory in (3+1)D was first computed in \cite{Buividovich:2008kq,Buividovich:2008gq} resulting in the extraction of the entropic $C$-function. Several lattice studies have been performed subsequently~\cite{Velytsky:2008sv,Itou:2015cyu,Rabenstein:2018bri,Rindlisbacher:2022bhe,Bulgarelli:2023ofi,Jokela:2023rba,Ebner:2024mee,Amorosso:2023fzt}. Computing the von Neumann entanglement entropy in Eq.~(\ref{eqn:VNEE}) is not feasible, so one computes instead the R\'{e}nyi entropy defined as 
\begin{equation}
    S^{(q)}=\frac{1}{1-q}\log\lp(\Tr\hat{\rho_V}^q\rp)\,,
\end{equation}
employing the replica trick, with $q$ denoting the number of replicas. 
The von Neumann entropy is defined as $S_E=\lim \limits_{q\to 1} S^{(q)}$, but the exact analytic continuation in $q$
is not feasible in general since only integer $q$ are possible on a lattice. 

The entanglement entropy in field theories typically takes form \cite{Buividovich:2008kq,Ryu:2006bv,Ryu:2006ef,Nishioka:2006gr,Klebanov:2007ws}
\begin{equation}
\label{eq:entropyUVfinite}
S=S_{UV}+S_f\,,
\end{equation}
where $S$ here denotes either the R\'{e}nyi or von Neumann entropy, $S_f$ represents the finite part of this entropy, and $S_{\rm UV}$ its ultraviolet divergent piece. In this work, we will demonstrate numerically for (2+1)D Yang-Mills theory that an entanglement entropy measure we define on the lattice (and henceforth call ``flux tube entanglement entropy" or FTE$^2$ in short), 
\begin{equation}
\label{eqn:RenyiDiff}
\tilde{S}^{(q)}_{\vert Q \bar{Q}} \equiv S^{(q)}_{\vert Q \bar{Q}}-S^{(q)} ,
\end{equation}
is finite in the continuum limit, with the UV divergent contribution canceling between the two terms on the r.h.s. The first term $S^{(q)}_{\vert Q \bar{Q}}$ is the Yang-Mills R\'{e}nyi entropy in presence of static quark sources defined as the gauge-invariant correlator of Polyakov lines and $S^{(q)}$ is the corresponding R\'{e}nyi entropy of the vacuum\footnote{A precise definition and discussion of the computation of this quantity is provided in Section~\ref{subsec:Polyakov}.}. This FTE$^2$ can be understood as the entanglement of a region of color flux $V$ in the chromoelectric flux tube formed between the heavy quark pair with the region $\bar V$ outside it after subtracting the corresponding quantity in the Yang-Mills vacuum. 

Before we discuss the properties of FTE$^2$, it is important to address a fundamental conceptual concern regarding the gauge invariance of entanglement entropy measures. This arises from the fact that the Hilbert space of gauge-invariant states cannot be expressed as the tensor product of two gauge-invariant subspaces: $\mcH \neq \mcH_\mcA \otimes \mcH_\mcAbar$.
Therefore the naive procedure of taking the partial trace over the complement region $\bar V$ to arrive at the  gauge-invariant reduced density matrix in Eq.~(\ref{eqn:rho_reduced}) is not suitable.
Fortunately, there are ways to define a gauge-invariant reduced density matrix on the lattice \cite{Buividovich:2008gq,Donnelly:2011hn,Casini:2013rba,Radicevic:2014kqa,Ghosh:2015iwa,Aoki:2015bsa,Soni:2016ogt}.  A fundamental observation in this regard is that due to the strongly ultraviolet divergence of entanglement entropy, it is not a property of states in Hilbert space alone but also of the algebra of observables  \cite{Witten:2018zxz}. We will offer our detailed perspective on this discussion (especially with regard to FTE$^2$) in Section \ref{sec:def_entent}. In brief, one approach is to extend the Hilbert space to include non-gauge invariant states (the extended Hilbert space approach) \cite{Buividovich:2008kq,Casini:2013rba} and the other is to analyze the algebra generated by gauge-invariant operators in regions $V$ and ${\bar V}$, which will contain the reduced density matrix $\hat{\rho_V} $\cite{Casini:2013rba,Radicevic:2014kqa}. 

In the extended Hilbert space approach, one considers a larger Hilbert space equal to the tensor product of partial Hilbert spaces defined over $V$ and $\bar{V}$ that are gauge-invariant everywhere but the boundary. 
To recover the original space, the extended Hilbert space is then restricted by requiring that the boundary flux of the states in $\bar{V}$ match those in $V$.
The original physical Hilbert space can then be written as a direct sum of the product of partial Hilbert spaces with matching flux.
The resulting density matrix is then gauge-invariant, a consequence of the orthogonality of spaces with different flux at the boundary.

In the algebraic approach, one writes the algebras $\mcA$ corresponding to the set of all gauge-invariant operators living completely in $V$, and likewise $\mcAbar$, those  living in $\bar V$, and finds the common center of gauge-invariant operators on the boundary. Diagonalizing the operators of this center, one obtains in the corresponding basis that the density matrix $\rho_{\mcA, \mcAbar}$ (the unique density matrix belonging to the product algebra $\mcA \mcAbar$) becomes block-diagonal, with each block corresponding to a superselection sector. One then takes the partial trace of each block, recovering a reduced density matrix which is block diagonal as well. This reduced density matrix $\hat{\rho}_\mcA$ is gauge-invariant by construction, and therefore, its associated entanglement entropy is gauge-invariant as well.

A source of ambiguity in the algebraic approach is that including or excluding states generated by operators on the boundary of $\mcA/\mcAbar$ in the domain of the reduced density matrix  has a large effect on the resulting entanglement entropy~\cite{Casini:2013rba}. 
The dependence of the entanglement entropy on these boundary operators causes the entanglement entropy to split into two terms, a ``boundary classical'' term and a ``bulk quantum'' term, with the former solely and explicitly dependent on the treatment of boundary operators. Small changes in the algebra $\mcA$ can dramatically affect the center elements, changing the number of blocks and the boundary classical term of the entropy.
It has been argued that the relative entropy and mutual information are both monotonic under inclusion of algebras, so different boundary operator inclusion conventions are required to converge in the continuum limit~\cite{Casini:2013rba}.  These two entropic measures are however difficult to realize on the lattice. For the entropic C-function computed with the extended lattice construction in \cite{Buividovich:2008gq}, all derivatives are evaluated using the electric center convention with the same area of the entangling surface; this cancels out the universal divergences allowing one to unambiguously extract information about the finite part of the entropy.

\begin{figure}[ht!]
\includegraphics[width=.35\textwidth]{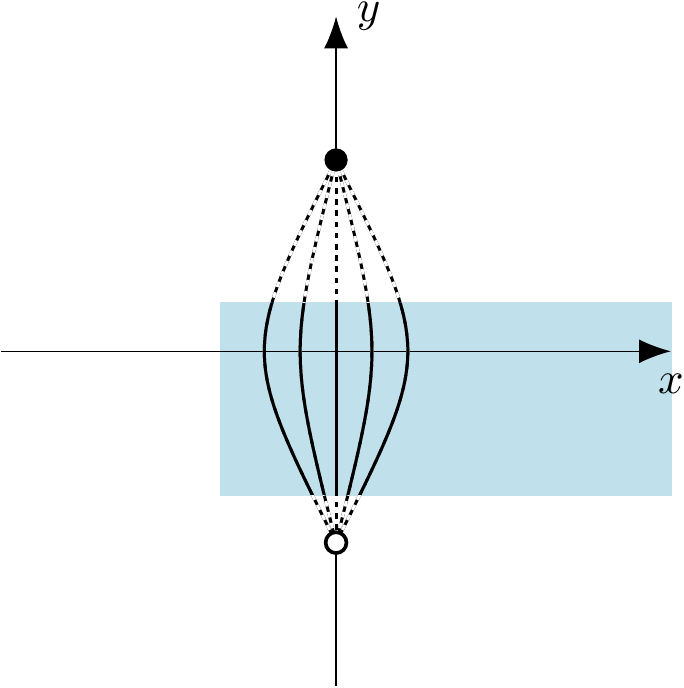}\\
\caption{\label{fig:crudehalfslab}
The ``half-slab" geometry used in this paper to calculate the Flux Tube Entanglement Entropy (FTE$^2$). The blue shaded region $V$ can be moved to fully cross-cut the flux tube or only partially. The dependence of FTE$^2$ on the width of the slab and its position relative to the quark-antiquark pair can also be studied systematically. }
\end{figure}

The requirement that the area of the entangling surface stay constant, however, is fairly restrictive.
While suitable for studying entanglement of the vacuum, the effect of degrees of freedom such as the size of region $V$ cannot be studied on the lattice with the entropic C-function. For FTE$^2$ instead, the UV contributions to the entanglement entropy cleanly cancel when subtracting the vacuum entropy, even when the area of the entangling surface is varied. Thus FTE$^2$ is not only manifestly gauge-invariant but finite as well. 
Further, with FTE$^2$, we can study the flux tube's excess entropy and its dependence on quark-antiquark separation, the dependence on the number of colors, the geometry of region $V$, its location, etc.

To study these dependencies, we will employ the geometry outlined in Figure \ref{fig:crudehalfslab}.
We make region $V$ a slab with specified width in the $y$ direction that spans half of the lattice in the $x$ direction. We can then move the slab $V$ to encompass the two extremes of it either completely separating the quark and the antiquark--cutting the flux tube--or not intercepting the flux tube at all.
Therefore, in moving the slab along the $x$ axis, we can explore gradually the process of cross-cutting the flux tube and its effect on FTE$^2$. Likewise, when we translate the slab in the $y$-direction, we can explore the variation of FTE$^2$ along the flux tube, studying whether the gluon fields nearer to quark/antiquark sources are more or less entangled with the rest of the string/bulk than those residing in the center. 
We can also in principle vary the width of the slab and its overall geometry, and for each of these configurations, further vary the number of colors, and the temperature, to study the dependence of $\tilde{S}^{(q)}_{\vert Q \bar{Q}}$ on these physical quantities.

The flexibility of FTE$^2$ allows us to construct models that can be compared to lattice data obtained for differing control parameters. As noted earlier, an EFT of long strings describes the excitations of a color flux tube on the lattice. We will employ here the simplest realization of this framework, the relativistic thin string model \cite{Luscher:1980iy}. In this model, the flux tube is a long thin string with minimal deflections. There are two sources of contributions to FTE$^2$ that one can address in this framework. One is the ``vibrational entropy'' due to transverse vibrational modes of the flux tube. The other is the ``internal entropy," the entropy due to the quantum states residing inside the flux tube that cannot be explained by vibrational dynamics alone. For the vibrational entropy, we discretize the string as a number of points determined by the UV cutoff, evenly spaced along the longitudinal axis. These points are free to move in the transverse direction and the string follows a linear interpolation between adjacent points.
This discretization of the string, combined with the Hamiltonian, allows us to solve for the entanglement entropy of different regions of the string partitioned along the longitudinal axis, providing a baseline expectation of the entropy when the string is fully cut.

To model the internal entropy of region $V$ of the color flux tube, we studied QCD in (1+1)D and  observed that FTE$^2$ is only dependent on the number of colors and whether this region intersects the flux tube or not. In particular, we are able to show that the internal entropy is directly proportional to the number of boundaries crossed by the string. This work will be published separately~\cite{FTEE_SUN2d:inprep}. For the (2+1)D $SU(2)$ Yang-Mills numerical study of interest here, the analytical (1+1)D results suggest that the numerical results will be sensitive to the probability of the string intersecting with region $V$, in particular, allowing one to  predict the dynamics of the FTE$^2$ when the flux tube is partially cut. 

The structure of the paper is as follows. In Section~\ref{sec:def_entent}, we summarize the discussion in the literature on how one defines entanglement entropy in gauge theories and offer our perspective on this discussion. We argue that the FTE$^2$ we propose should be  finite, and robust to local changes in the inclusion/exclusion of boundary operators. Specifically, we argue that FTE$^2$ will give the same result when computed with so-called trivial-center and electric center algebras. As we will discuss, it is not clear at present how to compute on the lattice algebras corresponding to a magnetic center.
In Section~\ref{subsec:Polyakov}, we define FTE$^2$ on the lattice as the product of  correlators of traces of Polyakov lines on different replicas. Since this quantity is manifestly gauge-invariant, it should follow that it should be independent of the choice of center algebra.  
The thin string model expectations and the analytical results of the (1+1)D Yang-Mills study are summarized in Section~\ref{sec:models}. The results of this study are presented in Section~\ref{sec:results}. We begin by providing the relevant details of the (2+1)D Yang-Mills Monte Carlo simulations in Section~\ref{sec:MC}. The results of our study are then presented in Section~\ref{sec:MCresults}. 
We first demonstrate that FTE$^2$ is finite in the continuum limit and therefore a robust measure of entanglement in the color flux tube. We then study its properties as a function of the half-slab geometry as the slab fully or partially cross-cuts the flux tube, or not at all. This includes as well the properties of FTE$^2$ as the width and location of the slab is varied along the flux tube. We find that our results point qualitatively to the presence both of the internal (color) and vibrational entropy and are consistent, respectively, with the expectations from our study of FTE$^2$ in (1+1)D Yang-Mills, and from the thin string model. More definitive quantitative statements will require an extension of our study to larger number of colors, more non-trivial slab geometries, and implementing refinements to the string model, in particular incorporating the intrinsic width of the string. We summarize our results in Section~\ref{sec:summary_outlook}, discuss the implications of these results, and further directions of research.  The computation of the entanglement entropy due to vibrational modes in the thin-string model is discussed in Appendix~\ref{app:thin_string}.

\section{Defining entanglement entropy in lattice gauge theory
  \label{sec:def_entent}}

In this section, we will provide an overview of definitions of the entanglement entropy in lattice gauge theory and present our perspective on their implications for the entanglement entropy of infrared (finite-size) objects. 
Our discussion follows closely the discussion in the papers cited in this section, in particular
Refs.~\cite{Casini:2013rba,Ghosh:2015iwa}, which should be consulted for a more in-depth perspective on the topic.
As noted earlier, the standard definition of entanglement entropy would not apply in a gauge theory because its
gauge-invariant Hilbert space $\mcH$ is not factorizable in terms of physical gauge-invariant Hilbert spaces  $\mcH_V$
and $\mcH_{\bar{V}}$ describing fields in regions $V$ and $\bar{V}$, respectively.
As a consequence, we cannot take the partial trace as in Eq.~(\ref{eqn:rho_reduced}) required to compute the entropy as in Eq.~(\ref{eqn:VNEE}).
This issue has been discussed at length in the literature
\cite{Buividovich:2008gq,Donnelly:2011hn,Casini:2013rba,Radicevic:2014kqa,Aoki:2015bsa,Ghosh:2015iwa,Soni:2015yga,Agarwal:2016cir}, 
and one can resolve it through alternative definitions of the reduced density matrix.
We will first discuss the algebraic approach to this problem and the associated ambiguities, and then discuss
the ``extended Hilbert space'' approach (currently implemented in lattice simulations) and its ambiguities due to gauge fixing.

\subsection{Center algebras and ambiguity in the entanglement entropy
\label{sec:EEdef_algebra}}

The most general approach to entanglement entropy in a gauge theory is to compute it from a reduced density matrix $\rho_V$ defined as a gauge-invariant operator that a), commutes with all fields in $\bar V$ and b), gives the same expectation values 
$\Tr(\rho_V \,\mc O_V)=\Tr(\rho\, \mc O_V)$ as the full density matrix $\rho$ for all gauge-invariant observables
$\mcO_V$ defined only on fields in $V$.
In more formal language, $\rho_V$ is an element of a gauge-invariant algebra of operators acting on $\mcH_V$ and is unique for any specific choice of algebra~\cite{Casini:2013rba}.
However, unlike in scalar field theories, there is no unique ``natural'' choice of algebra itself because of the local (gauge) symmetry.
Our discussion here will be based on (finite) Abelian groups, but is also  applicable to 
to continuous non-Abelian groups with some modifications.

Let $\Psi[\{U\}]$ be a complex wave functional on gauge link matrices,
and let $\mcG$ be the vector space of all $\Psi[\{U\}]$.
The subspace $\mcH\subset\mcG$ contains all gauge-invariant functionals satisfying 
$\Psi[\{U\}]=\Psi[\{U\}^g]$, with the links $\{U\}^g$ being the result of the gauge transformation $g$ of $\{U\}$.
The full algebra $\mcA(\mcG)$ of operators acting on $\mcG$ are generated by analogs of quantum operators of
coordinate and momentum, which we will call magnetic and electric operators, respectively, acting on each link $l$
independently~\cite{Casini:2013rba}. 
A magnetic generator $\hat{U}_l^R$ ``measures'' the link potential ${U}_l^R$ in the one-dimensional  representation
$R$, 
\begin{equation}
    \hat{U}_l^R\ket{\Psi[\{U_{1}\dots U_{l}\dots U_{N}\}]} ={U}_l^R\ket{\Psi[\{U_{1}\dots U_{l}\dots U_{N}\}]}\,,
\end{equation}
while an electric generator $\hat{L}_l^{g}$ translates the wave functional by shifting the link potential $U_l$ by group
element $g$,
\begin{equation}
    \hat{L}_l^{g}\ket{\Psi[\{U_{1}\dots U_{l}\dots U_{N}\}]} =\ket{\Psi[\{U_{1}\dots gU_{l}\dots U_{N}\}]}\,.
\end{equation}
Together, these operators generate the subalgebra of $\mcA(\mcG)$ associated with link $l$, with the 
electric and magnetic operators generating separate commuting subalgebras in the case of Abelian groups\footnote{
  In the non-Abelian case, electric operators do not commute only on the same link, generating a gauge group algebra instead.}.
For any (sub)set of links $X$, the algebra $\mcA(\mcG_X)\subset \mcA(\mcG)$ is given by 
the tensor product over $l\in X$.

The algebra $\mcA(\mcG)$ is not gauge-invariant. 
Gauge-transformation operators $\hat{T}_a^g \equiv \prod_b \hat{L}_{(ab)}^{(g)}$, where $(ab)$ is a link between
connected sites $a$ and $b$, commute with electric operators but not magnetic operators on the relevant links.
However the product of magnetic operators $\hat{U}^R_{(ab)}\hat{U}^R_{(bc)}$  commutes with
$\hat{T}_b^g$ 
and, as a result, the product of links along any closed (Wilson) loop $W^R_\Gamma\equiv \prod_{l\in\Gamma}U^R_l$ is invariant
under $\hat{T}_a^g$ for any site $a$. (In the non-Abelian case, a trace must also be taken.)
This operator modifies the state by adding a closed loop of electric flux~\cite{Radicevic:2014kqa,Aoki:2015bsa}.

\begin{figure}[ht!]
\includegraphics[width=.5\textwidth]{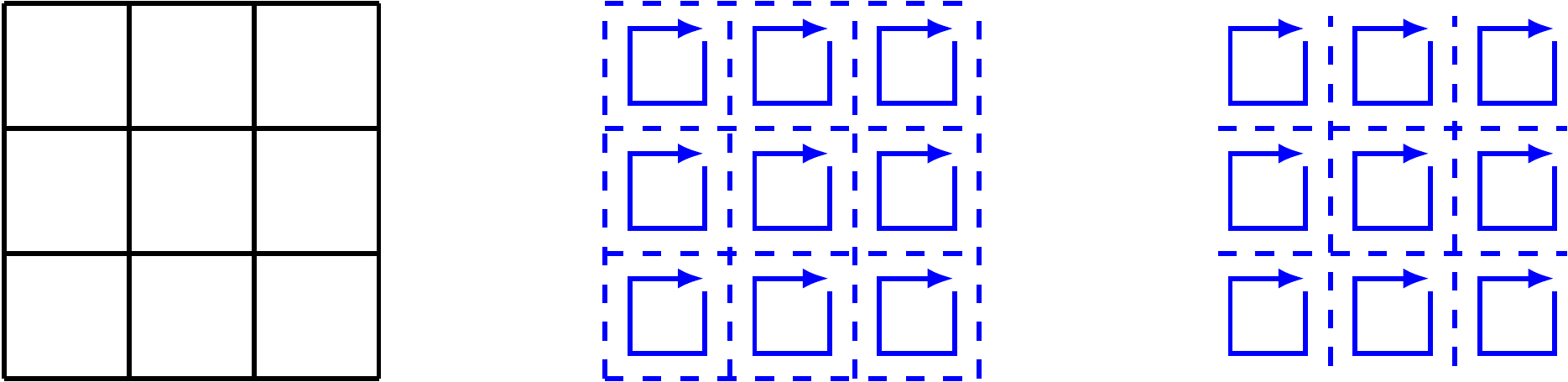}\\
\caption{\label{fig:ElecMagAlgebra}
A set of links $V$ (left) and its corresponding algebras $\mcA_\mcE$ (center), and $\mcA_\mcM$ (right). Dashed lines represent electric operators and solid loops represent Wilson loops. Note that the electric algebra contains boundary electric operators, while the magnetic algebra does not.}
\end{figure}

\begin{figure}[ht!]
\includegraphics[width=.5\textwidth]{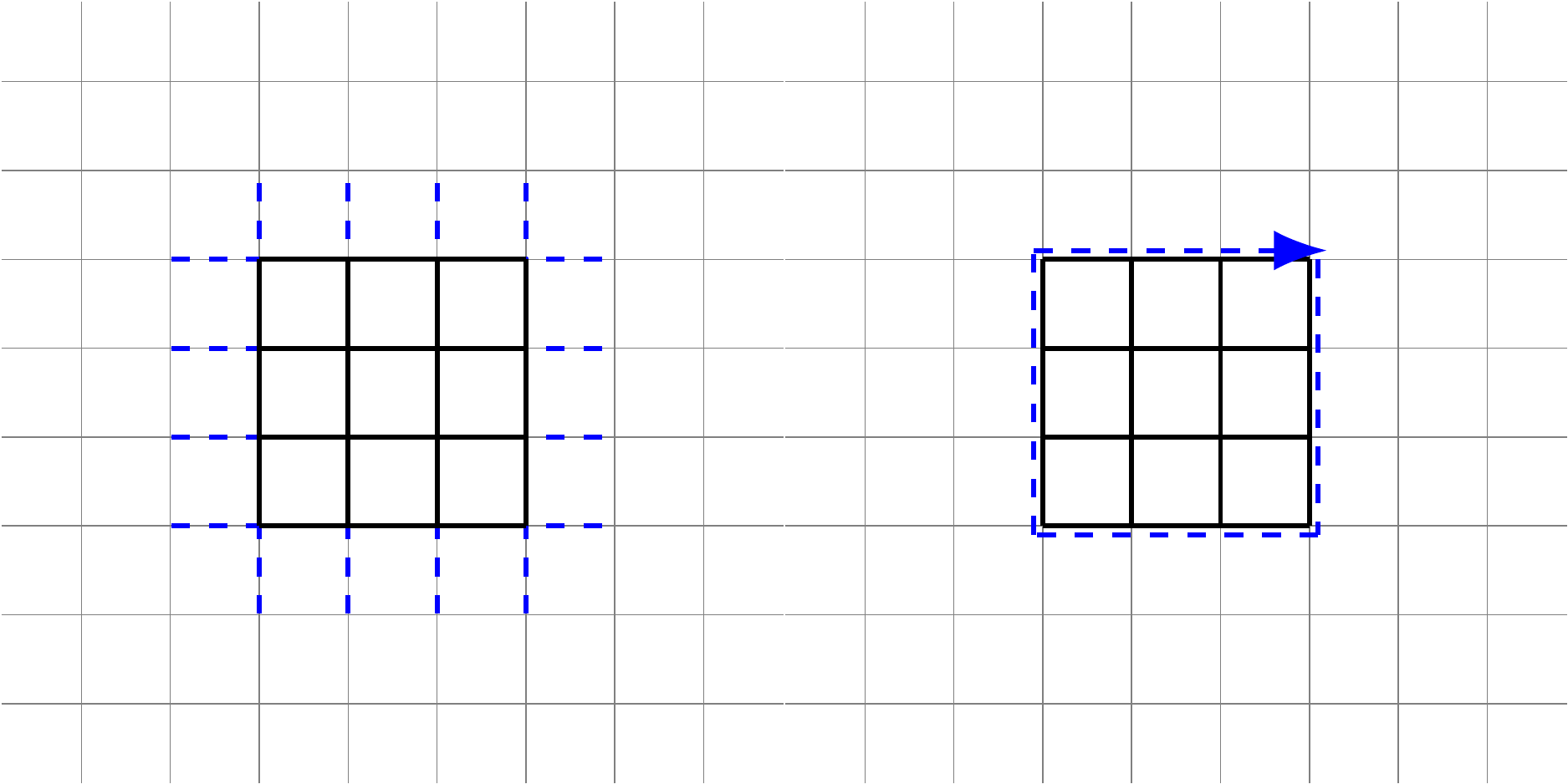}\\
\caption{\label{fig:ElecMagCenter}
Electric center (left, dashed lines) and magnetic center (right, dashed loop), shown for the set of links $V$ depicted with solid lines.}
\end{figure}

To generate a gauge-invariant algebra $\mcA_V$ associated with spatial region $V$,
only Wilson loops and electric operators on links in $V$ must be selected, 
up to the ambiguity of including or excluding operators on boundary links from the algebra.
Including all electric operators on the boundary yields the electric center algebra $\mcA_\mcE$, while excluding all of
them yields the magnetic-center algebra $\mcA_\mcM$.
These two algebras are depicted in Fig. \ref{fig:ElecMagAlgebra} and their resulting centers in Fig.
\ref{fig:ElecMagCenter}.
Another choice is to exclude specific boundary electric operators from the algebra to result in a trivial center.

\begin{figure}[ht!]
\includegraphics[width=.3\textwidth]{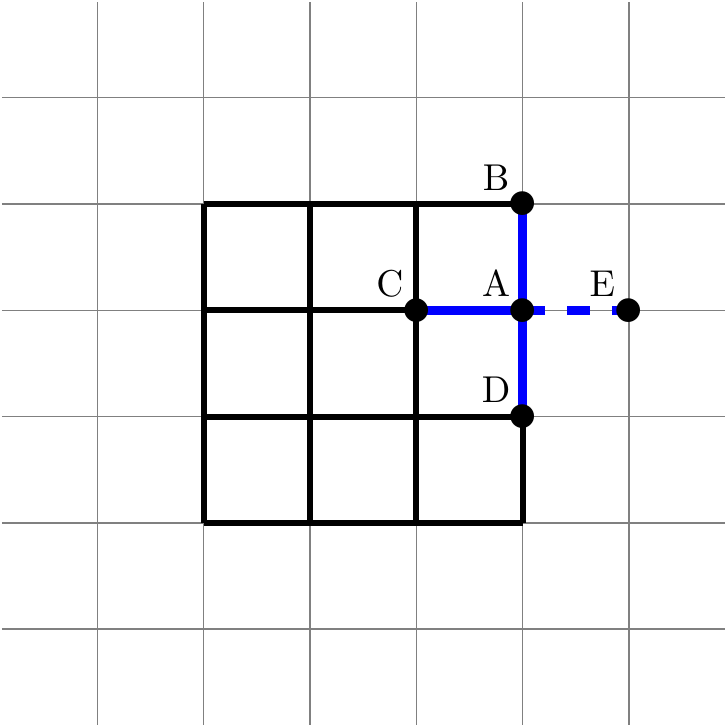}\\
\caption{\label{fig:GaugeOperator}
A gauge transformation at site $A$ affects four adjacent links and leaves the space $\mcH$ invariant.
The product of electric operators in region $V$ (the solid square) $L^{(AB)}_g L^{(AC)}_g L^{(AD)}_g$ 
is equivalent on $\mcH$ to $(L^{(AE)}_g)^{-1}$, an operator outside region $V$.
This operator commutes with all elements of the gauge-invariant algebra $\mcA_V$ and its commutant $\mcA'$; it is 
therefore an element of the center of $\mcA$, $\mcA'$, and combined algebra $\mcA\mcA'$.
}
\end{figure}

The algebra for the complementary set of links $\bar{V}$ is given by the commutant $\mcA'$ of the algebra $\mcA$ 
(the linear space of operators that commute with all elements of $\mcA$).
This region $\bar{V}$ may be separated from region $V$ by a ``buffer zone'' depending on the choice of algebra~\cite{Radicevic:2014kqa}.
Specifically, the gauge-invariant algebras for $V$ and $\bar V$ are the double commutants (the set of operators that commute with the commutant) of their gauge-invariant generators.
Thus the electric- and magnetic-center algebras are
\begin{align}
\label{eq:doublecomm}
\mcA_{\mc E} &= \{\hat{L}_l^{g},\hat{W}^R_\Gamma\}''_{l,\Gamma \in V}\,,
\\
\label{eq:doublecommMag}
\mcA_{\mc M} &= \{\hat{L}_l^{g},\hat{W}^R_\Gamma\}''_{l \in V \backslash \partial V,\, \Gamma \in V, }\,,
\end{align}
respectively, where $''$ denotes the  double commutant and $\partial V$ is the set of links with both ends on the boundary of $V$.
The choice of algebra, $\mcE$ or $\mcM$, will not matter for the general discussion below but we will return to their specific features later in the section.

For a general algebra choice ${\mc A}$, the combined algebra ${\mc A \mc A'}$ generated by $\mcA \cup \mcA'$ is not the full algebra of gauge-invariant operators: it excludes Wilson loops that intersect the boundary.
Further, it is not factorizable; it cannot be represented as a tensor product because $\mcA$, $\mcA'$, and $\mcA\mcA'$ share a center subalgebra $\mcZ$, which is a consequence of gauge invariance.
Indeed, the gauge-transformation operator $\hat{T}_g^a=\prod_b \hat{L}_g^{(ab)}$ maps a gauge-invariant state to itself,
\begin{equation}
    \hat{T}_g^a \ket{\Psi} = \ket{\Psi} \,,
\end{equation}
with $\big(\prod_b \hat{L}_g^{(ab)}\big)_\mcH = 1$ when restricted to the gauge-invariant subspace $\mcH$.
Thus for site $a$ on the boundary, the product of adjacent electric operators in $V$ can be expressed through electric operators in $\bar V$, as depicted in Fig.~\ref{fig:GaugeOperator}.
Such operators belong to both algebras\footnote{The magnetic algebra does not include electric operators on the boundary; its center elements are instead Wilson loops lying completely on the boundary. One can also in principle choose an algebra containing a set of boundary electric operators such that the algebra has no center elements aside from the identity. This is referred to as the trivial algebra/trivial center.} 
$\mcA$ and $\mcA'$ and therefore comprise the center $\mcZ=\mcA\cap\mcA'\in\mcA\mcA'$.

Once the center subalgebra is chosen, the partially-traced density matrix can be defined by first observing that all the center operators can be simultaneously diagonalized~\cite{Casini:2013rba}
\begin{equation} 
\label{eqn:centerblockdiagonal}
\Lambda = \left(\begin{array}{cccc}
[\lambda_\Lambda^1] & 0 & \hdots & 0 \\
0 & [\lambda_\Lambda^2] & \hdots & 0 \\
\vdots & \vdots & &\vdots \\
0 & 0 & \hdots & [\lambda_\Lambda^m]
\end{array}\right)\,,
\quad \forall\Lambda\in\mcZ_{\mcA\mcA'}
\end{equation}
where each block is proportional to an identity matrix.
Thus the center dictates how the gauge-invariant Hilbert space is divided into mutually orthogonal ``superselection sectors,'' invariant subspaces for operators in $\mcA\mcA'$.
Further, each superselection sector is now factorizable.
In the same basis, all operators in $\mcA\mcA'$ must be block-diagonal in order to commute with the center operators in Eq.~(\ref{eqn:centerblockdiagonal}).
Therefore, the global density matrix $\rho$ can also be restricted to its block-diagonal part~\cite{Casini:2013rba}
\begin{equation}
\label{eqn:DensMatBlockDiagonal}
\rho_{\mcA \mcA'}= 
\left(\begin{array}{cccc}
p_1 \rho_{\mcA_1 \mcA_1'} & 0 & \hdots & 0 \\
0 & p_2 \rho_{\mcA_2 \mcA_2'} & \hdots & 0 \\
\vdots & \vdots & &\vdots \\
0 & 0 & \hdots & p_m \rho_{\mcA_m \mcA_m'} 
\end{array}\right)\,,
\end{equation}
where each $\rho_{\mcA_k\mcA_k'}$ has unit trace and the probabilities $p_i$
reflect the probability distribution over states with definite values of all the center operators, $\sum_m p_k=1$.
Note that the expectation values are the same,
$\Tr\big[\rho_{\mcA\mcA'}\, \mcO_{\mcA\mcA'}\big] = \Tr\big[\rho\, \mcO_{\mcA\mcA'}\big]$,
for any operator $\mcO_{\mcA\mcA'} \in \mcA\mcA'$.

In other words, the off-block-diagonal elements of $\rho$ do not contribute to $\langle\mcO_{\mcA\mcA'}\rangle$ and can be dropped to get $\rho_{\mcA\mcA'}$, which is uniquely defined for the choice of $\mcA$ and preserves all physical expectation values, as required by definition.
Taking the partial trace of each block, one obtains the unique reduced density matrix $\hat{\rho}_\mcA$ \cite{Casini:2013rba},
\begin{equation}
\label{eqn:RedDensMatBlockDiagonal}
\hat{\rho}_{\cal A}= 
\left(\begin{array}{cccc}
p_1 \rho_{{\cal A}_1} & 0 & \hdots & 0 \\
0 & p_2 \rho_{{\cal A}_2} & \hdots & 0 \\
\vdots & \vdots & &\vdots \\
0 & 0 & \hdots & p_m \rho_{{\cal A}_m} 
\end{array}\right)\,,
\end{equation}
where each $\rho_{\mcA_k}$ has unit trace. 
Its von Neumann entropy
\begin{equation}
\label{eqn:EE_classical_shannon}
S_E=-\sum \limits_{k}p_k\text{log}p_k +\sum \limits_{k}p_kS(\rho_{\mcA_k}),
\end{equation}
has two terms, a ``boundary classical'' term $-\sum_{k}p_k\text{log}p_k$, and a ``bulk quantum'' term $\sum_{k}p_kS(\rho_{\mcA_k})$, with $S(\rho_{\mcA_k})$ the von Neumann entropy associated with density matrix $\rho_{\mcA_k}$~\cite{Casini:2013rba}.
The boundary classical term is the Shannon entropy of the distribution ${p_i}$ 
while the bulk quantum term is the average of the entanglement entropies corresponding to each block.
The Rényi entropy can be similarly expressed as 
\begin{equation}
\label{eqn:EE_classical_shannon_Renyi}
S^{(q)} = \frac1{q-1} \Big( - \log\sum_k p_k^q 
    -\log\sum_k a^{(q)}_k \text{Tr}(\rho_{\mcA_k}^q) \Big),
\end{equation}
where $a^{(q)}_k = p_k^q/(\sum_i p_i^q)$. 
The first term for both entropies is dependent on the center of the algebra; the second term in the Rényi entropy  has a form similar to its bulk counterpart in Eq.~(\ref{eqn:EE_classical_shannon}).

The boundary classical term alone makes the entanglement entropy center-dependent.
The generators of the electric center and the magnetic-center algebras in Eqs.~(\ref{eq:doublecomm}) and (\ref{eq:doublecommMag}), respectively, differ only by electric operators on the boundary.
Both $\mcA_\mcE$ and $\mcA_\mcM$ contain the same gauge-invariant operators in the bulk of region $V$ and, intuitively, all the corresponding physical observables should be independent of the boundary conventions in the continuum limit. However, as shown in Fig.~\ref{fig:ElecMagCenter}~(left), the electric center has a number of generators proportional to the size of the boundary,
while the magnetic center shown in Fig.~\ref{fig:ElecMagCenter}~(right) has only one generator: the Wilson loop surrounding the boundary.
This difference can lead to a large center-dependent discrepancy in the boundary classical term~\cite{Casini:2013rba,Bulgarelli:2024onj} due to ultraviolet degrees of freedom.
In particular, this has been shown to be the case for the wave function of a topological $Z_2$ model~\cite{Buividovich:2008gq,Casini:2013rba}. 

On the other hand, for a state describing only an infrared-scale object, the choice of algebra should not matter in the continuum limit ($a\to0$). 
Consider a wave functional with support only over long-wavelength ($\lambda\gg a$) field configurations and their gauge transformations.
Removing boundary electric operators must be equivalent to contracting region $V$ by a distance of at most $O(a)$, and doing so cannot change the number of effective degrees of freedom in the continuum limit.
Therefore, if there is a measure of entanglement with finite continuum-limit value for such a state, it should not depend on the choice of details of the algebra on the boundary.

We expect this to be the case for FTE$^2$ defined in Eq.~(\ref{eqn:RenyiDiff}) if the UV and IR contributions to it decouple as in Eq.~(\ref{eq:entropyUVfinite}).
This is the conjecture we will explore in this paper by computing FTE$^2$ with the electric center algebra construction. In this context, adding the color flux between $Q$ and $\Qbar$ to the vacuum state changes the distribution $p_k$ over superselection sectors and, likely, the entropies $S(\rho_{\mcA_k})$ within each of the sectors. The cancelation of the UV-divergent parts of Eq.~(\ref{eqn:EE_classical_shannon_Renyi}) in Eq.~(\ref{eqn:RenyiDiff}), with and without the extra color flux, is therefore a non-trivial result and we will demonstrate it later in Section~\ref{sec:results}.

For non-Abelian continuous gauge groups, the procedure to obtain the entanglement entropy is similar.
The magnetic center can still be formed from Wilson loops on the boundary.
The electric center elements become quadratic Casimir operators at boundary vertices, but this change does not affect the form the entanglement entropy takes in Eq.~(\ref{eqn:EE_classical_shannon}) ~\cite{Lin:2018bud,Soni:2015yga}.
In the electric center case, the superselection sectors of Eq.~(\ref{eqn:centerblockdiagonal}) correspond to different irreducible representations of non-Abelian electric flux on the boundary transforming under the gauge group~\cite{Donnelly:2011hn}.
In the non-Abelian case, the irreducible representations are not one-dimensional.
As a result, if one takes the trace over states in $\mathcal{G}$ instead of $\mathcal{H}$ to arrive at Eq.~(\ref{eqn:RedDensMatBlockDiagonal}), the entanglement entropy receives a contribution  $\sum_{l}\langle\log(d_{R,l})\rangle$, where $d_{R,l}$ is the dimension of the representation $R$ of flux on boundary link $l$ ~\cite{Lin:2018bud}.
This additional term can be seen when the extended Hilbert space approach is applied to non-Abelian gauge theories.

\subsection{Extended Hilbert space and the electric center algebra
\label{sec:EEdef_Hext_U1}}
In this subsection, we will discuss the extended Hilbert space approach to defining the entanglement entropy.
This construction matches the electric center algebra choice, and it has been implemented on a Euclidean lattice using the replica trick~\cite{Buividovich:2008kq,Aoki:2015bsa,Ghosh:2015iwa}.
Let us examine the issue with factorizing a global wave functional 
$\Psi[\{U\}_V, \{U\}_{\bar V}] = \Psi[\{U\}^g_V, \{U\}^g_{\bar V}]$,
which is invariant with respect to gauge transformation $g$.
The product of gauge-invariant spaces $\mcH_V\otimes\mcH_{\bar V}$ does not cover the entire Hilbert space $\mcH$.
To cover $\mcH$, one has to construct the ``extended'' Hilbert space
$\mcH_\text{ext}=\mcH'_V\otimes\mcH'_{\bar V}$~\cite{Buividovich:2008gq,Casini:2013rba,Radicevic:2014kqa},
where the partial Hilbert spaces $\mcH'_V$, $\mcH'_{\bar V}$ are gauge-invariant everywhere
except\footnote{
  The boundary $\partial V$ between \emph{sets of links} $V$ and $\bar V$ is a \emph{set of sites} adjacent to at
least one link from each $V$ and $\bar V$.} on the boundary $\partial V$.
Such an extension is necessary because the global Hilbert state $\mcH$ contains states with electric flux through the
boundary~\cite{Buividovich:2008gq}.
Any state from $\mcH_\text{ext}$ can then be projected onto its gauge-invariant subspace $\mcH\subset\mcH_\text{ext}$.

To demonstrate this\footnote{
  In this formal discussion, we do not consider splitting boundary links in half as in
  \cite{Buividovich:2008gq,Casini:2013rba}.
  Instead, we ``draw" the boundary through sites in $\partial V$ by assigning some of the adjacent links to $V$ and
  some to $\bar V$ as in \cite{Radicevic:2014kqa}; this does not impact any of our conclusions.}, consider a contribution to the wavefunctional $\Psi$
\begin{equation}
\Psi[\{U\ldots\}_V, \{U\}_{\bar V}] 
  = \int \prod_{s\in\partial V} dg_s \, \Psi_{V} [\{U\}^{g_s}_V] \, \Psi_{\bar V}[\{U\}^{g_s}_{\bar V}]\,,
\end{equation}
where the partial wave functionals $\Psi_{V}$ and $\Psi_{\bar V}$ belong to $\mcH'_V$ and $\mcH'_{\bar V}$, respectively,
but the integral (Haar group-average) over transformations on the boundary ensures that $\Psi$ is 
gauge-invariant on the boundary as well.
This convolution over the gauge degrees of freedom enforces  Gauss' law (conservation of electric flux) on the boundary.
To demonstrate this at a single point of the boundary, consider a transformation at site $A$ (see Fig.~\ref{fig:GaugeOperator})
affecting four links
\begin{equation}
\Psi[U_{AB},U_{AC},U_{AD},U_{AE},\ldots]
  = \int dg_A \Psi_{V} [g_A U_{AB}, g_A U_{AC}, g_A U_{AD},\ldots] \, \Psi_{\bar V}[g_A U_{AE},\ldots]\,,
\end{equation}
where $,\ldots$ in the arguments denote the rest of the links in $V$ or $\bar V$.
The convolution over $g_A$ can be rewritten as a sum over definite integer values of electric flux $\lambda_A$
going from $\bar V$ to $V$ through site $A$, and it is an eigenvalue of the electric center element (discussed in  Sec.~\ref{sec:EEdef_algebra}) associated with site $A$.

In the case of the $U(1)$ gauge group with $U=e^{i\theta}$, $g=e^{i\alpha}$, and $\int dg=\int (d\alpha/2\pi)$, 
\begin{equation}
\label{eqn:psi_sum_SSsectors}
\Psi[U_{AB},U_{AC},U_{AD},U_{AE},\ldots] 
= \sum_{\lambda_A=-\infty}^{\infty} 
\Psi_{V,\lambda_A} [U_{AB},U_{AC},U_{AD},\ldots] \Psi_{\bar V,-\lambda_A}[U_{AE},\ldots]\,,
\end{equation}
where $\Psi_{V,\lambda_A}$ is defined to be 
\begin{equation}
\label{eqn:psi_proj_SSsectors}
\Psi_{V,\lambda_A} = \hat P_{V,\lambda_A} \Psi_{V} 
  = \int_0^{2\pi}\frac{d \alpha}{2\pi} e^{i\lambda_A \alpha}
    \Psi_{V}[e^{i\alpha} U_{AB}, e^{i\alpha} U_{AC}, e^{i\alpha} U_{AD},\ldots]\,,
\end{equation}
with $\lambda_A$ an eigenvalue of the total electric field $E_l=-\dot \theta_l$ on the links from site $A$ into $V$:
\begin{equation}
(\hat E_A)_V \Psi_{V,\lambda_A} 
  = i\Big[\frac{\partial}{\partial\theta_{AB}} + \frac{\partial}{\partial\theta_{AC}} 
    + \frac{\partial}{\partial\theta_{AD}}\Big]\Psi_{V,\lambda_A}
  = \lambda_A \Psi_{V,\lambda_A}\,,
\end{equation}
and similarly for $\Psi_{\bar V,-\lambda_A}[U_{AE},\ldots] = P_{\bar V,-\lambda_A}\Psi_{\bar V}$.
Note that these eigenvectors are \emph{gauge-covariant}\footnote{
  It should be noted that $\Psi_{V,\lambda_A}$ cannot be in superposition with $\Psi_{V,\lambda_A'}$ from a different sector. 
  This would violate gauge invariance, just as would a superposition of states with different total electric charge.
  Such a constraint can only be imposed in the language of algebras but not the Hilbert space $\mcH_V$ itself~\cite{Lin:2018bud}.
  The projection in Eq.~(\ref{eqn:psi_proj_SSsectors}) should be thought of as a transformation to the electric-flux bases in $\mcH'_V$ and $\mcH'_{\bar V}$ that can be factors of the factorizable component $\Psi_{\lambda_A} = \hat P_{V,\lambda_A} \hat P_{\bar V,-\lambda_A}\Psi$ of the global wave function in a specific superselection sector $\lambda_V$.
} with respect to the transformation 
$g_A=e^{i\alpha_A}$ on site $A$,
\begin{equation}
\label{eqn:wf_cov_proj}
\Psi_{V,\lambda_A}[\{U\}_V^{g_A}] = e^{-i\lambda_A\alpha_A} \Psi_{V,\lambda_A}[\{U\}_V]\,,\quad
\Psi_{\bar V,-\lambda_A}[\{U\}_{\bar V}^{g_A}] = e^{+i\lambda_A\alpha_A} \Psi_{\bar V,-\lambda_A}[\{U\}_{\bar V}]\,,
\end{equation}
but their transformations mutually cancel in Eq.~(\ref{eqn:psi_sum_SSsectors}) making the full wavefunctional gauge-invariant. The gauge covariance of the wave functionals indicates that the fields in state $\Psi_{V,\lambda_A}$ are compatible with 
having charge 
$+\lambda_A$ and in state $\Psi_{\bar V,-\lambda_A}$ with charge $-\lambda_A$ at site $A$ on the boundary, 
which is in transparent agreement with total electric flux $\lambda_A$ going from $\bar V$ to $V$ through site $A$.
These charges are not new degrees of freedom: they simply label different mutually orthogonal eigenspaces of
$(E_A)_{V(\bar V)}$ in $\mcH'_{V(\bar V)}$~\cite{Aoki:2015bsa,Ghosh:2015iwa}.

Applying this transformation at all points of the boundary $\partial V$, one arrives at the decomposition of the
global gauge-invariant space $\mcH$ over superselection sectors~\cite{Casini:2013rba,Ghosh:2015iwa},
\begin{equation}
\mcH = \bigoplus_{\{\lambda\}_{\partial V}} \mcH'_{V, {\{\lambda\}}} \otimes \mcH'_{\bar V,{\{-\lambda\}}}
\end{equation}
each of which is factorizable and allows one now to define a reduced density matrix
\begin{equation}
\label{eqn:rho_reduced_Hext}
\rho_V = \sum_{\{\lambda\}_{\partial V}}\Tr_{\mcH'_{\bar V,\{\lambda\}}}
    \Big[\hat P^{\{\lambda\}} |\Psi\rangle\langle\Psi|\Big]\,,
\end{equation}
where $\hat P_{\{\lambda\}} = \hat P_{V,\{\lambda\}} \hat P_{\bar V,\{-\lambda\}}$ 
The key to the definition in Eq.~(\ref{eqn:rho_reduced_Hext}) being gauge-invariant is that the extended Hilbert spaces with different
surface charge configurations are orthogonal. Hence $\rho_V$ does not contain matrix elements between different
superselection sectors and therefore satisfies the general argument for gauge invariance we discussed previously - see the discussion leading to Eq.~(\ref{eqn:RedDensMatBlockDiagonal}) in Sec.~\ref{sec:EEdef_algebra}.
The construction can be applied to non-Abelian gauge theory where one must sum now 
over all possible irreducible representations $R$ of the color group at each point of the
boundary~\cite{Donnelly:2008vx,Donnelly:2011hn}.
The transformed wave functions in Eq.~(\ref{eqn:wf_cov_proj}) become $\dim R$-multiplets $d_R$ from invariant subspaces of the
color-electric operator, resulting in contributions $\log(d_R)$ to the entanglement entropy~\cite{Donnelly:2014gva}.

The extended Hilbert space definition of the entanglement entropy can be implemented with the replica trick as a path integral on the lattice.
If all the boundary links are free integration variables, the entanglement entropy for the electric center algebra is recovered~\cite{Aoki:2015bsa}.
On the other hand, if the gauge is fixed on a maximal tree of boundary links, the entanglement entropy of the trivial-center algebra is obtained~\cite{Casini:2013rba}.
Therefore if an entanglement measure depends only on gauge-invariant observables \emph{away from the boundary}, as the FTE$^2$ we will define explicitly below in Eq.~(\ref{eqn:RenyiDiffPolyakov}), it will be unaffected by gauge fixing of boundary links\footnote{Other examples of center independent observables suggested in the literature include relative entropy and mutual information~\cite{Casini:2013rba,Berta:2014vma}. }.

 \section{Computation of the flux tube entanglement entropy (FTE$^2$)
  \label{sec:lat_comp_ftee} }

In this section, we will  construct FTE$^2$ explicitly and discuss its implementation on the lattice. 
Unlike the formal discussion in Section~\ref{sec:def_entent}, the Monte Carlo evaluation of the partition function does not provide a way to explore superselection sectors and split the entropy into bulk quantum $+$ boundary classical terms since the density matrix is only implicit in this framework.
While there is ambiguity in selecting the algebra for region $V$ in order to define its enganglement entropy, only the electric algebra has been implemented on the lattice~\cite{Buividovich:2008kq,Buividovich:2008gq,Itou:2015cyu,Rabenstein:2018bri}.
We will use the same lattice construction to study FTE$^2$, specifically to compute correlators of Polyakov lines that play the role of static quark sources $Q$ and $\bar Q$.
We will then study its dependence on the shape of region $V$ by ``carving out'' different fragments of the flux tube for the ``half-slab'' geometry shown in Fig.~\ref{fig:crudehalfslab}.

\subsection{FTE$^2$ from Polyakov line correlators
  \label{subsec:Polyakov}}

The formal expression for the Euclidean density matrix for gauge fields at temperature $1/L_t$ is
\begin{equation}
\label{eqn:lat_denmat}
\langle U^\text{out} | \rho | U^\text{in}\rangle
  = \frac1{Z} 
    \int _{U(t=0) = U^\text{in}} ^{U(t=L_t) = U^\text{out}} \mcD U  
     e^{-S_g[U]}\,,
\end{equation}
and it naturally yields the reduced density matrix by tracing over (making periodic) the fields in the complement region $\bar V$~\cite{Buividovich:2008kq}:
\begin{equation}
\label{eqn:lat_denmat_reduced}
\langle U_V^\text{out} | \rho_V | U_V^\text{in}\rangle
  = \frac1{Z} 
    \int _{U_V(t=0) = U_V^\text{in}} ^{U_V(t=L_t) = U_V^\text{out}} \mcD U_V  
    \int_{U_{\bar{V}}(t=0) = U_{\bar{V}}(t=L_t)} \mcD U_{\bar{V}}
    \,\, e^{-S_g[U_V, U_{\bar{V}}]}\,,
\end{equation}
where $Z=\int_{V\oplus\bar{V}} \mcD U e^{-S}$
is the partition function to normalize the trace of the density matrix to one,
\begin{equation}
\int \mcD U_V \,\langle U_V | \rho_V | U_V \rangle = 1\,.
\end{equation}
This density matrix corresponds to the electric-center algebra~\cite{Aoki:2015bsa}.

In order to create a color flux tube, we insert static quark and antiquark sources at sites in region $\bar V$.
This is equivalent to adding Polyakov loops to the partition function and the density matrix.
Indeed, the static heavy quark Hamiltonian contains only the temporal hopping term, ${\mathcal H}_Q = (\sum_{\vec x} Q^\dag_{\vec x} U_{\vec x,\hat t}^\dag Q_{\vec x} + \text{h.c.})$. (Our discussion also applies equivalently to antiquarks.)
For a single quark at point $\vec y_Q$ with initial and final color states $i_Q,i_Q'$, the density matrix is the product of temporal links
\begin{equation}
\label{eqn:lat_denmat_reduced_Q}
\langle U,\, \vec y_Q^\prime, i_Q^\prime |\rho_V|U,\, \vec y_Q,i_Q\rangle
  \propto \delta_{y_Q^\prime,y_Q} \, 
   {\Big[\Big\langle \prod_{\tau=0}^{L_t-1} U_{\hat t} (\vec y_Q,\tau)\Big\rangle\Big]^{i_Q^\prime  i_Q}}\,,
\end{equation}
averaged with the same distribution as the pure Yang-Mills density matrix defined in Eq.~(\ref{eqn:lat_denmat_reduced}).
In this work, we restrict ourselves to placing the quarks and antiquarks in the complement region $\bar{V}$. Their color degrees of freedom are therefore traced over resulting in a temporal (Polyakov) loop $P_{\vec x} = \Tr\prod_{\tau=0}^{L_t-1} U_{\hat t} (\vec x,\tau)$ for each quark (and a complex conjugate loop for each antiquark).
The normalized reduced density matrix is obtained by dividing this quantity by a complete trace of the density matrix, which yields a regular correlator of Polyakov loops at the same spatial points:
\begin{equation}
\label{eqn:lat_denmat_reduced_Qtr}
\langle U_V^\text{out}| \rho_{V|Q_{\vec x}, \Qbar_{\vec y},\ldots} |U_V^\text{in}\rangle
  = \frac
   {\big\langle P_{\vec x} \, P^\dag_{\vec y} \, \dotsm \big\rangle_{U_V^\text{out}; U_V^\text{in}}}
   {\big\langle P_{\vec x} \, P^\dag_{\vec y} \, \dotsm \big\rangle_{\phantom{U_V^\text{out}; U_V^\text{in}}}}\,.
\end{equation}
The denominator is the usual correlator of Polyakov loops on a periodic lattice. 
The numerator is a correlator of identical Polyakov loops, with the only difference that the gauge fields in region $V$ are fixed to their initial and final values as in Eq.~(\ref{eqn:lat_denmat_reduced}):
\begin{equation}
\label{eqn:lat_qqbar_single_open}
\langle U_V^\text{out} | \rho_V | U_V^\text{in}\rangle
  = \frac1{Z} 
    \int _{U_V(t=0) = U_V^\text{in}} ^{U_V(t=L_t) = U_V^\text{out}} \mcD U_V  
    \int_{U_{\bar{V}}(t=0) = U_{\bar{V}}(t=L_t)} \mcD U_{\bar{V}}
    \,\, e^{-S_g[U_V, U_{\bar V}]}\,  \big[ P_{\vec x}P^\dag_{\vec y}\dotsm\big],
\end{equation}

\begin{figure}[ht!]
  \centering
\includegraphics[width=.4\textwidth]{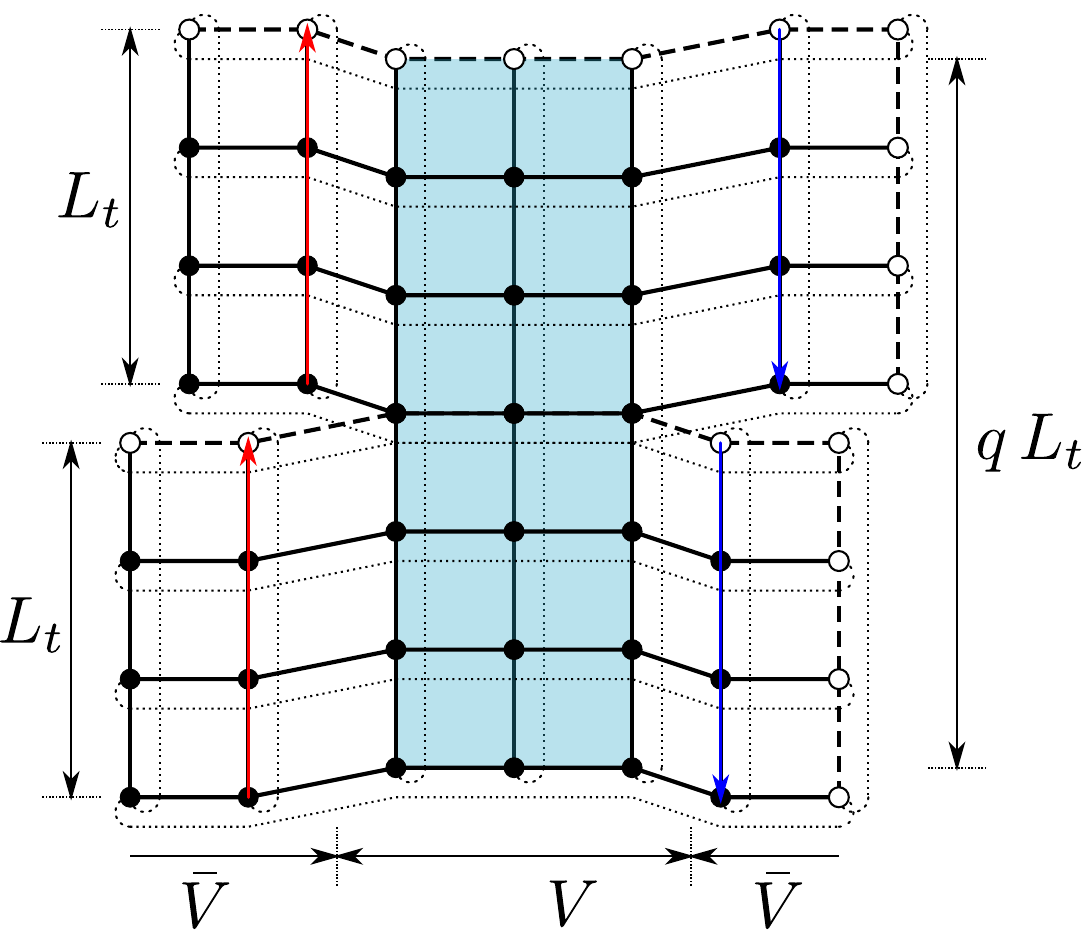}
\caption{\label{fig:pants}
Correlator of Polyakov loops representing a static quark (red) and an antiquark (blue) on the $q=2$ replicas of an $L_x\times L_\tau=6\times3$ lattice.
The spatial boundary conditions are $L_x$-periodic. 
The temporal boundary conditions are $qL_t$-periodic in region $V$ (shaded) and $L_t$-periodic in region $\bar V$.
Open points on the spatial and temporal edges are wrap-around images, and the dashed lines indicate copies of the same site.
}
\end{figure}

As described in \cite{Callan:1994py,Calabrese:2004eu}, the power $q$ of the reduced density matrix, $[\hat\rho_V]^q$, is obtained by stacking $q$ independent replicas of the path integral in Eq.~(\ref{eqn:lat_denmat_reduced}) in the temporal direction.
For the links in region $\bar{V}$, the temporal boundary conditions are 
$U_\mu^{(r)}(L_t)=U_\mu^{(r)}(0)$ for each replica $r=1\ldots q$,
implementing the trace in the functional space over the fields.
The link variables in region $V$ are identified for consecutive replicas $r$ and $r+1$ as $U_\mu^{(r)}(L_t)=U_\mu^{(r+1)}(0)$, thus implementing the product in the functional space. 
Also, the $qL_t$-periodic boundary conditions $U_\mu^{(q)}(L_t)=U_\mu^{(1)}(0)$ then implement the overall trace of $(\hat\rho_V)^q$, for a $Q\Qbar$ pair in $\bar V$ resulting in
\begin{equation}
\Tr[(\hat\rho_V)^q] 
  = \frac{Z^{(q)}_{|Q\bar Q}}{[Z_{|Q\bar Q}]^q}
  = \frac{\big\langle \prod_r^q P^{(r)}_{\vec x} \, P^{(r)\dag}_{\vec y}\big\rangle}
   {\big[\big\langle P_{\vec x} \, P^\dag_{\vec y}\big\rangle\big]^q}
   \,\cdot\, \frac{Z^{(q)}}{Z^q}
   \,,
\end{equation}
where $Z^{(q)}_{|Q\bar Q}$ and $Z_{|Q\bar Q}$ are partition functions in presence of a quark-antiquark pair on $q$-replica and ordinary lattices.
Note that the factor $(Z^{(q)}/Z^q)$ appears in the r.h.s. because the $q$-replica correlator of Polyakov loops is normalized by $Z^{(q)}$, the partition function of Yang-Mills fields on the $q$-replica lattice, while the partial correlator in Eq.~(\ref{eqn:lat_qqbar_single_open}) is normalized by $Z$, the partition function on the  ordinary lattice.
The geometry of a lattice with $q=2$ replicas is shown in Fig.~\ref{fig:pants}.

We can now formally define the FTE$^2$ that we introduced previously in Eq.~(\ref{eqn:RenyiDiff}), and express it directly from the ratio of Polyakov loop correlators:
\begin{align}
\label{eqn:RenyiDiffPolyakovprequel}
\tilde{S}^{(q)}_{\vert Q \bar{Q}} = S^{(q)}_{\vert Q \bar{Q}} -S^{(q)} 
  &= -\frac{1}{q-1} \left(\log
  \frac{Z^{(q)}_{\vert Q \bar{Q}}}{Z^{q}_{\vert Q \bar{Q}}}-\log
  \frac{Z^{(q)}}{Z^{q}}\right)
\\
\label{eqn:RenyiDiffPolyakov}
  &= -\frac{1}{q-1} \log
  \frac{\langle \prod\limits_{r=1}^{q}\Tr P_0^{(r)}\Tr P_{\vec{x}}^{(r)\dag}\rangle}
       {\big[\langle\Tr P_0\Tr P_{\vec{x}}^\dag\rangle\big]^q}
\,.
\end{align}
In other words, FTE$^2$ is given by correlation of Polyakov loops across replicas, normalized by their correlation within each replica.
The advantage of this construction is that one does not have to evaluate both terms on the r.h.s of Eq.~(\ref{eqn:RenyiDiffPolyakovprequel}) separately, which is substantially harder on the lattice. 

In our construction, replicas are joined at time slices of spatial links by a fixed-time slice of spatial links.
In principle, one could place this ``cusp", namely, the $D-2$ dimensional boundary region separating both replicas and regions $V$ and $\bar{V}$, inside  a timelike plaquette instead of on a vertex. This is for instance what is done in \cite{Chen:2015kfa}.
It has been speculated \cite{Bulgarelli:2024onj} that the center plaquette construction of \cite{Chen:2015kfa} could correspond to an entanglement entropy defined on algebras with a different center.
In a separate publication in preparation~\cite{FTEE_SUN2d:inprep}, we demonstrate explicitly that FTE$^2$ in (1+1)D Yang-Mills does not depend on whether the cusps are placed on the vertices or in the centers of the plaquettes.

\subsection{Color and vibrational contributions to FTE$^2$
  \label{sec:models}}
To approximate the entropy of a segment of the color flux tube, one may consider two sources of quantum entanglement.
Firstly, the color-electric flux inside the flux tube should give rise to entanglement because its color state must match
at the $V/\bar V$ boundary but may change along the flux tube, for example, due to longitudinally propagating gluons.
Further, the shape of the flux tube should also contribute to the entanglement entropy because it must be continuous across the $V/\bar V$ boundary.
We therefore conjecture that FTE$^2$ can be partitioned as 
\begin{equation}
\label{eq:vib+int}
\tilde{S}^{(q)}_{\vert Q \bar{Q}} = S_\text{internal} + S_\text{vibrational}\,,
\end{equation}
where $S_\text{internal}$ is the entanglement entropy associated with internal color degrees of freedom, while $S_\text{vibrational}$ is the entanglement entropy associated with the orientation and transverse fluctuations of the flux tube.
We neglect the intrinsic width of the flux tube~\cite{Caselle:2012rp} in this approximation.

The internal entanglement entropy should reflect the possible color states of the flux tube. 
For a flux tube carrying fundamental-representation $SU(N_c)$ electric flux, it is natural to expect that its $S_\text{internal}$ receives contribution $\propto\log N_c$ every time the flux tube crosses the boundary between $V$ and $\bar V$.
This is indeed the case in (1+1)-dimensional Yang-Mills theory, where a flux tube does not have transverse degrees of freedom and which is straightforward to compute.
In a paper in preparation~\cite{FTEE_SUN2d:inprep}, we show that FTE$^2$ in (1+1)-dimensional
Yang-Mills theory depends only on the number of colors, the dimension of the color source representation, and the number  of boundaries $F$ between $V$ and $\bar V$ crossed by the flux tube:
\begin{equation}
\label{eqn:1+1analytical}
\big(\tilde{S}^{(q)}_{\vert Q \bar{Q}}\big)_{\rm (1+1) YM}= F\cdot\log (N_c)\,.
\end{equation}
For example\footnote{
  As noted in \cite{Donnelly:2011hn}, the Yang-Mills vacuum entanglement entropy gets an additional contribution $\sum_R p(R) \log(\mathrm{dim}R)$ compared to Abelian theories, where $p(R)$ is the probability distribution over representations $R$ of the electric fluxes at each point of the boundary.
  This is compatible with the explicit result~(\ref{eqn:1+1analytical}), since the flux tube between $Q$ and $\bar Q$ ``biases'' the color-electric flux though the boundary.
}, $\tilde{S}^{(q)}_{\vert Q \bar{Q}}=2\log N_c$ for two fundamental-representation static quarks in the region $\bar V$ separated by a single ``slab'' of region $V$.
We have also shown in \cite{FTEE_SUN2d:inprep} that the result in Eq.~(\ref{eqn:1+1analytical})) is largely independent of lattice details, specifically of whether the replica and $V/\bar V$ boundaries are drawn through the corners or the centers of lattice plaquettes. 
These (1+1)-dimensional analytic results further motivate employing FTE$^2$ as a useful measure of quantum entanglement in higher dimensions.

The internal entanglement entropy of the flux tube should be independent of the distance between the
quark and the antiquark as long as the flux tube cannot ``avoid" crossing the spatial $V/\bar V$ boundaries.
However if there are additional (transverse) dimensions, the flux tube can fluctuate and avoid region $V$ (which is possible only if the quark and antiquark are in a spatially connected part of $\bar V$), or perhaps cross the boundary multiple times.
In this case, the internal entanglement entropy $S_{\rm internal}$ should be
\begin{equation}
\label{eq:intersectionProbability}
S_\text{internal}=\langle F\rangle \cdot \log(N_c)\,,
\end{equation}
where $\langle F\rangle$ is the average number of times the flux tube crosses the $V/\bar V$ spatial boundary.
Therefore, $S_\text{internal}$ in more than one spatial dimensions will also, in principle, depend on quark and
antiquark locations, as well as the geometry of the $V/\bar V$ boundary and the dynamical shape of the flux tube.
Since the probability distribution of small deflections of the flux tube is Gaussian,
we anticipate $\langle F \rangle$ to be described approximately by the error function of $x$, where $x$ is the spatial dimension on the (2+1)D lattice, as shown in  Fig.~\ref{fig:crudehalfslab}.
This interpretation will be explored in Section~\ref{sec:results}.

To study the vibrational entanglement entropy in Eq.~(\ref{eq:vib+int}), one may describe the color flux tube as a quantum vibrating string.
The effective theory of long strings has been extensively studied~\cite{Luscher:1980iy,Polchinski:1991ax,Luscher:2004ib,Drummond:2004yp,Aharony:2009gg,Athenodorou:2007du}. 
In this initial work, we compare our lattice results only to the predictions of the simplest model, and leave comparisons with state-of-the-art models for the future.
We briefly summarize this effective string model for the vibrational entanglement entropy below and for more details we refer readers to Appendix~\ref{app:thin_string}.
Since this model takes into account only transverse motion of the string, it depends only on the length of the string and the number of effectively independent oscillators, but not on $N_c$ or the representation of the color sources.
Discrepancies between this model and actual lattice results should be attributable to the internal (color) dynamics of the flux tube and can be studied by changing $N_c$ and the representation of the $Q\bar Q$ pair, which is also left for future work.

In the simplest model considered here, small transverse fluctuations of the ``thin string" Gaussian are described by the approximate Hamiltonian~\cite{Luscher:1980iy}: 
\begin{equation}
\label{eqn:Hamiltonian_thinstring}
H=M^2L+\frac{\pi}{2M^2L}\int\limits_{0}^{\pi} ds\, (p^2+M^4x'(s)^2)\,,
\end{equation}
where $x(s)$ is the displacement parameterized by the coordinate along the string $0 < s < \pi$, $M^2$ is the string tension, and $L$ is its length.
The endpoints of the string are fixed, $x(0)=x(\pi)=0$.
The potential depends on the string ``slope'' $x' = \frac{dx}{ds}$, and the kinetic energy on the momentum $p(s) \equiv -i\frac{\delta}{\delta x(s)}$.
There is only only one transverse direction in (2+1) dimensions, but in (d+1)-dimensional space-time each of the (d-1) transverse directions is governed by an independent copy of the Hamiltonian~(\ref{eqn:Hamiltonian_thinstring}), which would give the same contribution to the vibrational entanglement entropy.

\begin{figure}[ht!]
    \centering
    \includegraphics[width=.49\textwidth]{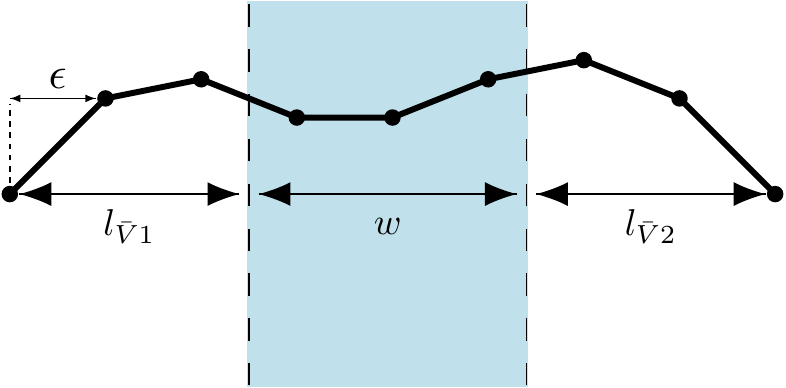}
    \caption{\label{fig:stringModelSchematic}
Discretized model of the thin string traversing region $V$ (shaded blue). 
The points are separated by distance $\epsilon$ along the longitudinal axis of the string.
Here $w$, $l_{\bar{V}1}$, and $l_{\bar{V}2}$ are, respectively, the segments of the string in region $V$, in region $\bar{V}$ to the left and to the right of $V$.}
\end{figure}

In its pure ground state, the string is described by a Gaussian density matrix of coupled harmonic oscillators~\cite{Bombelli:1986rw,Callan:1994py}.
Traced over the transverse fluctuations in the complement region $\bar V$, it yields a reduced Gaussian density matrix, from which the von Neumann and Rényi entanglement entropies can be computed.
Since the string has internal thickness (albeit neglected in this model), it cannot vibrate with wavelength shorter than some $\lambda$~\cite{Luscher:1980iy}.
It is important to note that $\lambda$ has physical meaning as the size of the string segment  representing an independent effective degree of freedom, and is different from the UV cutoff $a^{-1}$ on the lattice with spacing $a$.
In practice, we represent the string as a set of points separated by distance $\epsilon\sim\lambda$ along the longitudinal axis of the string.
These points can move in the transverse directions, and the shape of the string is linearly interpolated between them as shown in Fig. \ref{fig:stringModelSchematic}.

\begin{figure}[ht!]
  \centering
  \includegraphics[width=.49\textwidth]{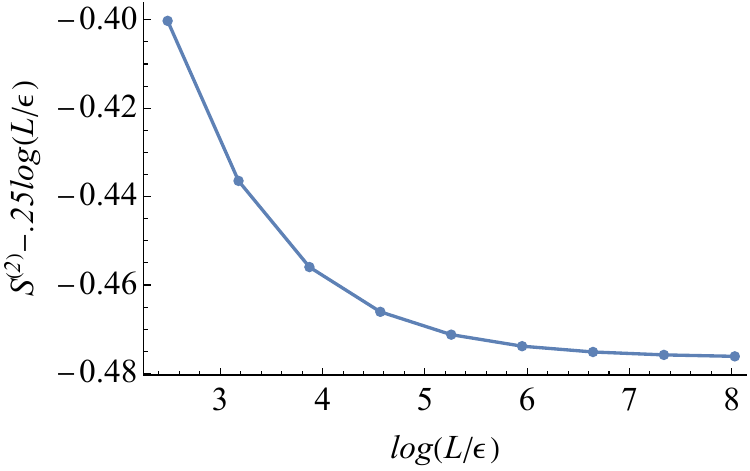}
  \includegraphics[width=.49\textwidth]{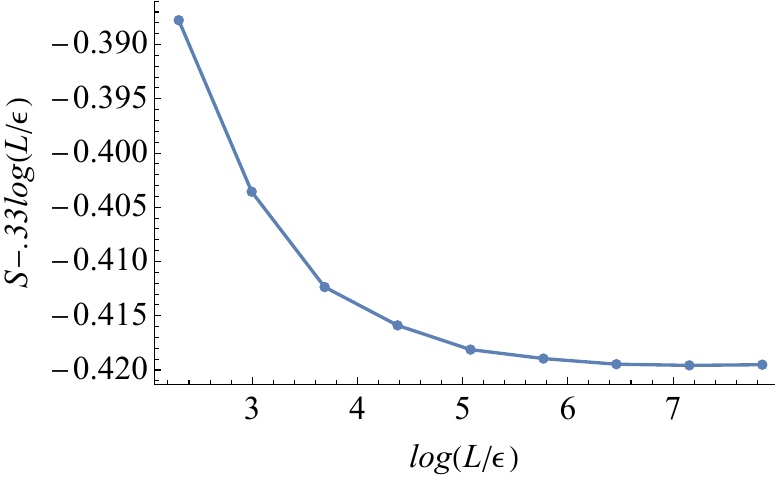}\\
  \caption{\label{fig:vnanalyticscaling}
    The finite part of the vibrational entanglement entropy of a thin string, cross-cut in the middle ($s/\pi=1/2$)
    by region $V$ of width $L/3$, for the $q=2$ Rényi entropy (left)
    and for the von Neumann entropy (right). This quantity asymptotes to a constant at large $L/\epsilon$, as anticipated from conformal field theory. 
}
\end{figure}

The entanglement entropy of the thin string model in the $L/\epsilon\to\infty$ limit exhibits behavior similar to free massless scalar theory and conformal field theory.
We find the leading logarithmic behavior $\frac{1}{3}\text{log}(L/\epsilon)$ for the von Neumann entanglement entropy and $\frac{1}{4}\log(L/\epsilon)$ for the Rényi entropy with $q=2$.
This behavior is consistent with the entanglement entropy in (1+1)D conformal field theories with periodic boundary conditions and central charge $c=1$~\cite{Calabrese:2004eu,Bazavov:2017hzi,Calabrese:2009qy}.
The emergent scaling behavior of these length scaling terms is seen in Fig. \ref{fig:vnanalyticscaling} (left) for the Rényi vibrational entropy and in Fig. \ref{fig:vnanalyticscaling} (right) for the von Neumann vibrational entropy for large values of $L/\epsilon$.

\begin{figure}[ht!]
  \centering
  \includegraphics[width=.49\textwidth]{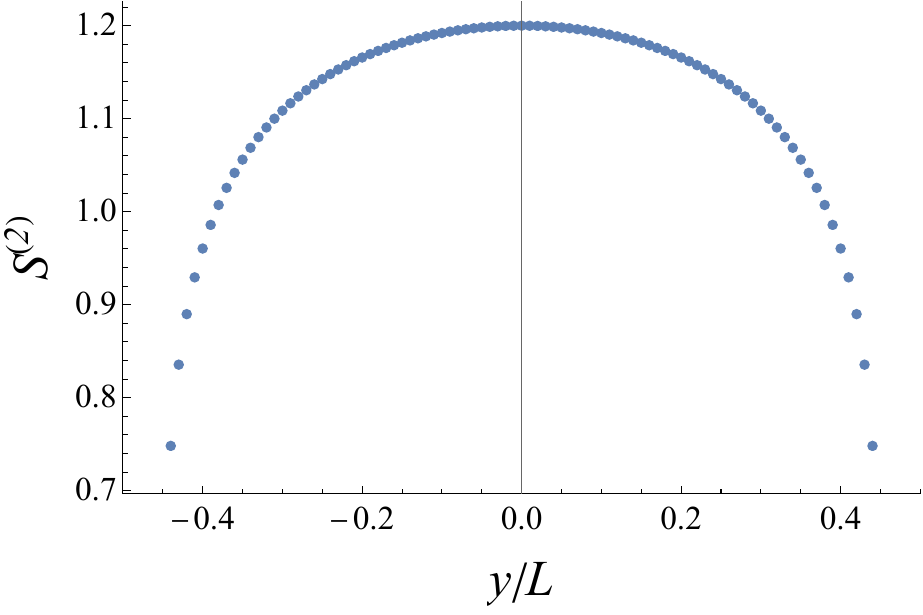}
  \includegraphics[width=.49\textwidth]{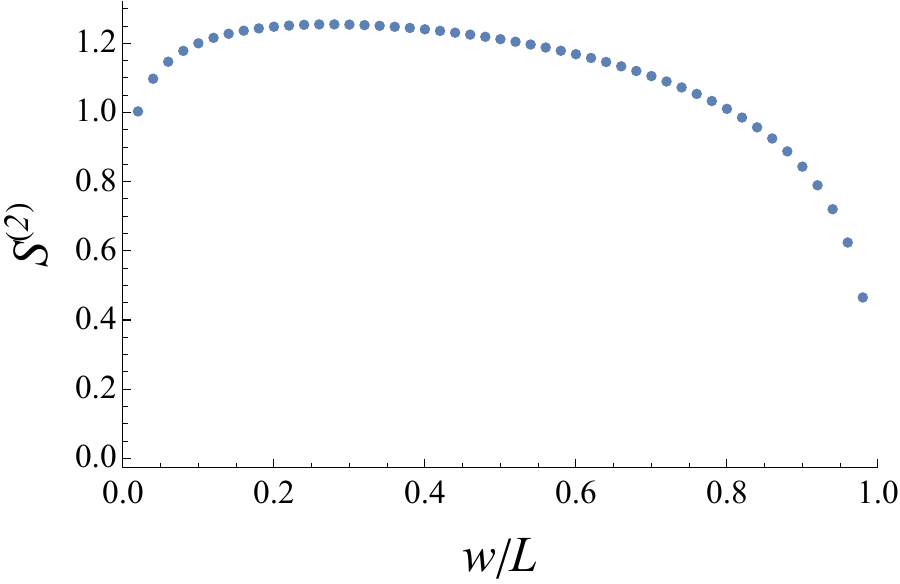}\\
  \caption{\label{fig:wLocAnalyticScaling}
    The dependence of the finite part of the Renyi vibrational entropy $S^{(2)}$ of the thin string on the 
    location (left) and the width (right) of region $V$ ($L/\epsilon=1000$ in the plots).
    Left: The Rényi entropy as a function of the longitudinal coordinate $y/L=s/\pi-0.5$ of the midpoint of region $V$.
    Region $V$ has width $\frac{w}{L}=1/10$.
    Right: The Rényi entropy as a function of $w$, the width of region $V$ placed in the middle of the string ($y=0$).
}
\end{figure}

We further explore the thin string model to find the dependence of \FTEE on how the string is partitioned into region $V$ and the complement regions $\bar V_1$ and $\bar V_2$ containing the color sources $Q$ and $\bar Q$.
This dependence converges and becomes independent of $L/\epsilon$, but is hard to parameterize as a function of longitudinal fractions $w/L$ and $l_{\bar V\,1,2}/L$.
We plot the Rényi entropy as a function of the location and the width of region $V$ in Fig. \ref{fig:wLocAnalyticScaling}.
The vibrational entropy exhibits mild dependence on the width and location of region $V$, with $y=0$ and $w\sim L/3$ configurations having the highest entanglement entropy.
In \cite{Calabrese:2004eu,Calabrese:2009qy}, the Rényi entanglement entropy of an interval of length $w$ within a periodic system of length $L$ in (1+1)D conformal field theory was found to have the string portion term $\frac{c}{6}(q-\frac{1}{q})\log(\sin(\frac{\pi w}{L}))$.
We recover this result within our model when we relax our boundary conditions from $x(0)=x(\pi)=0$ to $x(0)=x(\pi)$.

The model's dependence on longitudinal fractions may not accurately describe flux tubes where $\lambda$ is on the same order as $L$.
Taking $L/\epsilon\to\infty$ in our model allows us to shrink the domain of our basis states, making the microscopic local details of our string model largely irrelevant.
The effective string describing the color flux tube is not a set of points linearly linked as in Fig. \ref{fig:stringModelSchematic}, but as both $L/\lambda$ and $L/\epsilon$ become large, we expect their behavior to converge.
When $\lambda\sim L$, our degrees of freedom are not microscopic, and we can no longer make this argument.
Consider a string with no microscopic details- only global degrees of freedom.
This can be accomplished by 
constraining the string to be analytic (continuous derivatives of all orders),
as information contained in the entire string could then be ascertained from the transverse displacement at a single point and its derivatives to all orders.
In this case, the entanglement entropy of the string would still have a logarithmic dependence on the number of vibrational modes $L/\lambda$, but would lose its string portion term, as all partial traces one could take of the string's density matrix would leave one with a reduced density matrix representing the same mixed state ensemble.

To accurately reproduce the full entropic behavior we see on the lattice in (2+1)D that we will report on in Section~\ref{sec:results}, one will likely require refinements of the thin string model~\cite{Polyakov:1986cs,Braaten:1986bz}.
While we do not expect full agreement of the (2+1)D Yang-Mills lattice data with our prediction of the internal entropy in Eq.~(\ref{eq:intersectionProbability}), and of the above discussed string vibrational entropy
predictions, it is informative to test the degree to which they accurately describe the flux tube.
Given the flexibility of FTE$^2$, and with our ability to study its dependence on the width of the slab, its location in the flux tube, and the length of the flux tube in our lattice setup, we can test these predictions and the assumptions they are based on individually and find to which order they are accurate.

\section{2+1 D Yang-Mills simulations of FTE$^2$
  \label{sec:results}}
In this section, we present our calculations of \FTEE in $SU(2)$ Yang-Mills gauge theory on a (2+1)D lattice. 
We will, a) study the continuum limit of the \FTEE and check that it converges to a finite value, and b) compare  \FTEE to the expectations from the internal and the vibrational degrees of freedom discussed in Sec.~\ref{sec:models}.
For the former, we will decrease the lattice spacing $a$ while keeping the size $w$ of region $V$ and the length of the string $L$ fixed in physical units determined by the string tension.
For the latter, we will vary the size and the geometry the system at one selected lattice spacing.
We begin by providing the details of our Monte Carlo simulations, followed by the results, and comparisons to the expectations from the vibrational (thin string model) entropy and the internal (color) entropy.

\subsection{Details of Monte Carlo simulations}
\label{sec:MC}
\begin{figure}
\centering
\includegraphics[width=.49\textwidth]{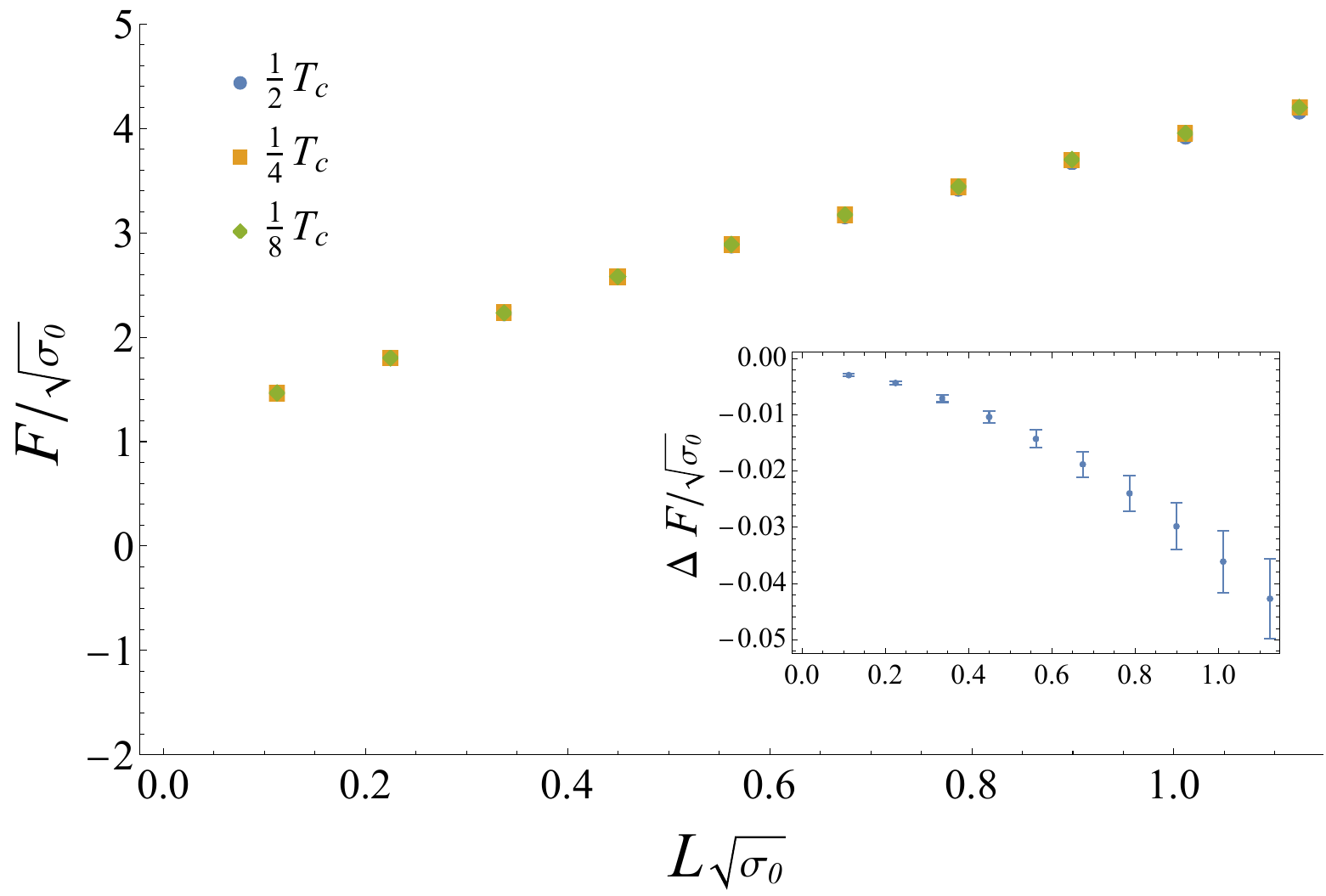}
\caption{\label{fig:FreeEnergyTemp} 
The free energy of a quark-antiquark pair at $\beta=12.630$ as a function of distance for different temperatures. 
The agreement between $T_c/2$ and the lower temperature results indicates that at $T=T_c/2$ the system is still dominated by the ground state.
The inset zooms in on the free energy difference between $T_c/2$ and  $T_c/8$. The string tension has a slightly lower value at $T_c/2$ relative to $T_c/8$.
}
\end{figure}

The Monte Carlo calculations in this work are performed in the confining regime of the (2+1)D Yang-Mills theory. 
To reduce statistical noise, we set the temperature to $T=T_c/2$,
where our $Q\bar Q$ system is expected to be dominated by the ground state.
Indeed, as shown in Fig.~\ref{fig:FreeEnergyTemp}, the free energies $F_{Q\bar Q}(T, r)$ of a $Q\bar Q$ pair calculated at $T=T_c/2$, $T_c/4$, and $T_c/8$ for a range of separations $r$ used in this work are in very close agreement. This is fortuitous because calculations at a lower temperature would require substantially more statistics due to the increased statistical noise in longer Polyakov loops.

For $T_c/2$, we set our time extent $L_t$ in lattice units and choose $\beta=\beta_c(L_t/2)$ as the critical coupling corresponding to a lattice with half of our time extent.
The values of the critical coupling $\beta$ corresponding to time extent $L_t/2$ are chosen following \cite{Billo:1996pu,Edwards:2009qw} and are listed in Table~\ref{tab:lattices_runs}.
The lattice data and parametrization given in \cite{Edwards:2009qw} were used to find the critical coupling for $\beta<20$ and the fit\footnote{
 Specifically, the first two $\beta$ values in Table~\ref{tab:lattices_runs} were taken from \cite{Edwards:2009qw} and $\beta=18.679$ from their parametrization $\beta_c(N_t)=1.5028(21)N_t+ 0.705(21)-\frac{0.718(49)}{N_t}$.
 The rest of the $\beta$ values were found using the parametrization $\beta_c(N_t)=N_c^2\big(0.380(3)N_t + 0.106(11)\big)$ in \cite{Billo:1996pu}.
 } 
 given in \cite{Billo:1996pu} for $\beta>20$.
As a cross-check, we extract the string tension from the static quark potential determined from Polyakov loop correlators computed away from region $V$, where they behave as if computed on a single-replica lattice.
We fit the static quark potential to a linear plus $(1/L)$ form with and without the Luscher term $(-\pi/(24L))$ fixed to extract the string tension.
The difference in the values extracted from each fit is used as the systematic uncertainty, which is combined with the statistical error to give the overall uncertainty.
We compare the values of the string tension $a\sqrt{\sigma(T_c/2)}$ extracted from our lattices at $T=T_c/2$ to the values of $a\sqrt{\sigma_0}=a\sqrt{\sigma(T=0)}$ from the fit in \cite{Teper:1998te},
\begin{equation}
\label{eqn:sigma0_fit}
a\sqrt{\sigma_0}=1.337(23)/\beta+0.95(38)/\beta^2+1.1(1.3)/\beta^3\,,
\end{equation}
and find that they are $\approx93-94\%$ of the $\sigma_0$ values, which is expected as the string tension decreases with temperature.
The latter are used to set the physical scale of our lattices and region $V$, see, for example, the width $w\sqrt{\sigma_0}$ in Table~\ref{tab:lattices_runs}.
While Fig. \ref{fig:FreeEnergyTemp} shows close agreement between the free energy at $T_c/2$ and $T_c/4$ (or $T_c/8$) at $\beta=12.630$, the values of the string tension for this value of $\beta$ are slightly different, $a\sqrt{\sigma(T_c/4)}=0.1109(17)$ and $a\sqrt{\sigma(T_c/8)}=0.1109(16)$, both of which agree within uncertainty with the value given by Eq.~(\ref{eqn:sigma0_fit}) in \cite{Teper:1998te}.

\begin{table}
\centering
\caption{\label{tab:lattices_runs}
  The details of Monte Carlo calculations of \FTEE in $SU(2)$ Yang-Mills theory on (2+1)-dimensional lattices $L_x \times L_y\times L_t$ with $2$ replicas:
  the gauge coupling $\beta = 4/(ag^2)$, $a\sqrt{\sigma}$ calculated from our lattice at $T=T_c/2$, the width $w$ of the region $V$ in lattice units, 
  and the number of (top-level) Polyakov-loop correlator samples $N_\text{samples}$.
  The fits to the potential (linear + fixed $\pi/(24L)$ L\"uscher term) are performed with uncorrelated $\chi^2$ for the two coarsest lattices and with $10^{-2}$ cut-off on the correlation matrix eigenvalues for the others.
  }
\begin{tabular}
{r|r|r|r|r r|r|r|r}
\hline\hline
$L_{x}\times L_{y}$ & $L_t$ & $\beta$ 
& $a\sqrt{\sigma}_{0}$ \cite{Teper:1998te} 
& $a\sqrt{\sigma}$ &[$\chi^2$/dof]
&$N_\text{samples}$ 
& $w/a$ & $w\sqrt{\sigma_0}$
\\
\hline
$64\times 16$ & 8 & 6.536 & 0.2307(106) 
& 0.2144(4) & [9.46/1]
& 4362 & 1 & 0.231(11)
\\
$128\times 32$ & 16 & 12.630 & 0.1124(31) 
& 0.1046(2) & [0.22/1]
& 1363 & 2 & 0.225(6)
\\
\multicolumn{6}{c|}{\textit{(same)}}
& 358 & 1 & 0.112(3)
\\
\multicolumn{6}{c|}{\textit{(same)}}
& 2646 & 3 & 0.337(9)
\\
$192\times 48$ & 24 & 18.679 & 0.0745(17) 
& 0.0695(4) & [0.08/1]
& 571 & 3 & 0.223(5)
\\
$256\times 64$ & 32 & 24.744 & 0.0557(11) 
& 0.0525(2) & [0.05/1]
& 1472 & 4 & 0.223(5)
\\
$320\times 80$ & 40 & 30.824 & 0.0444(8) 
& 0.0417(3) & [0.003/1]
& 1165 & 5 & 0.222(4)
\\
$384\times 96$ & 48 & 36.904 & 0.0369(7) 
& 0.0345(3) & [0.17/1]
& 679 & 6 & 0.222(4)
\\\hline\hline
\end{tabular}
\end{table}

To compute the Polyakov loop correlators, we use the standard Wilson plaquette $SU(2)$ Yang-Mills action, 
\begin{equation}
S = -\frac{\beta}{2} \sum_{x,\mu<\nu} \Re\Tr U_{x,\mu\nu}\,,
\end{equation}
performing sweeps with alternating
Kennedy-Pendleton heatbath updates~\cite{Kennedy:1985nu} and over-relaxation updates~\cite{Brown:1987rra}, and also incorporating the multilevel algorithm~\cite{Luscher:2001up}. 
Each sweep consists of one heatbath update and five overrelaxation updates of the entire lattice alternating between even- and odd-checkerboard sublattices.
The initial gauge configuration is thermalized with 15,000 global sweeps.
In the multilevel calculation of the Polyakov loop correlators, we use a 2-level algorithm.
The first sublevel updates are performed on a single replica, while the lowest-level updates are performed on temporal slabs of size $\frac{1}{2} T_c^{-1}$, which corresponds to 4 slabs per replica at $T=T_c/2$.
At the lowest level, we accumulate 1000 samples separated by one local sweep and augmented with the multi-hit procedure.
The top-level samples are separated by 100 global sweeps.
We analyzed our data with jackknife resampling and binned the data to monitor for autocorrelation effects.
We found that the flux tube has a wide profile, so the $x$ dimensions of our lattices are extended by a factor of $4$ to minimize wrap-around effects and ensure that the $x\to\mp\infty$ limits can be effectively reached on our lattices to realize the case where the half-slab does not completely cross-cut the flux tube.

As mentioned before, the region $V$ of the half-slab geometry is a rectangle of size $(L_x/2)\times w$, where the long side of the lattice $L_x\sqrt{\sigma_0}\approx14.3$.
We compute the Polyakov-line correlators at all values along the $x$ direction of the lattice.
The Polyakov loop correlator in the numerator of Eq.~(\ref{eqn:RenyiDiffPolyakov}) for the fully cross-cut \FTEE is averaged over points in the $(L_x/4)$ segment centered on the half-slab, which are at least $|\Delta x|\sqrt{\sigma_0}\geq1.8$ from the edge of the slab.
The denominator in Eq.~(\ref{eqn:RenyiDiffPolyakov}) requires computing Polyakov loop correlators on a single-replica lattice.
Instead, we compute these correlators on the same lattices by similarly averaging over points at $|\Delta x|\geq (L_x/8)$ away the half-slab,
to save on simulation cost and to eliminate correlated stochastic fluctuations.
\begin{figure}[ht!]
\includegraphics[width=.4\textwidth]{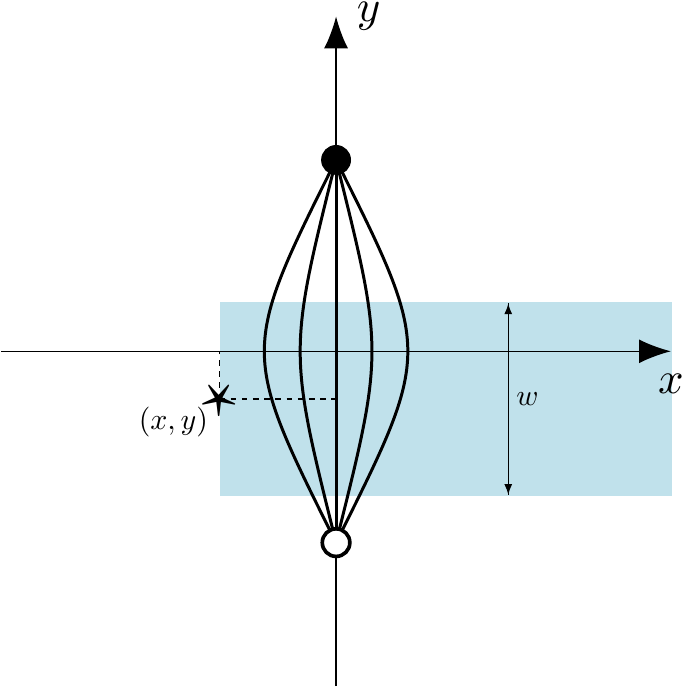}\\
\caption{\label{fig:refinedhalfslab}
The Flux Tube Entanglement Entropy (FTE$^2$) is studied using the ``half-slab" region $V$ (shaded blue) that has width $w$ in the $y$ direction and extends indefinitely to $x\to\infty$ (on a finite periodic lattice, the slab spans $L_x/2$ of the lattice).
We explore the dependence of FTE$^2$ on the width of the slab and its position relative to the quark-antiquark pair at $(0,\pm L/2)$.
Depending on the position of its left-most edge center $(x,y)$, the flux tube  centered at $(0,0)$ can be completely ($x\to -\infty$) or partially ($x\sim 0$) cross-cut by region $V$.
The $x\to+\infty$ limit determines the ``baseline'' entanglement entropy of the vacuum itself partitioned into $V$ and $\bar V$.
}
\end{figure}
\begin{figure}[ht!]
\centering
\includegraphics[width=.49\textwidth]{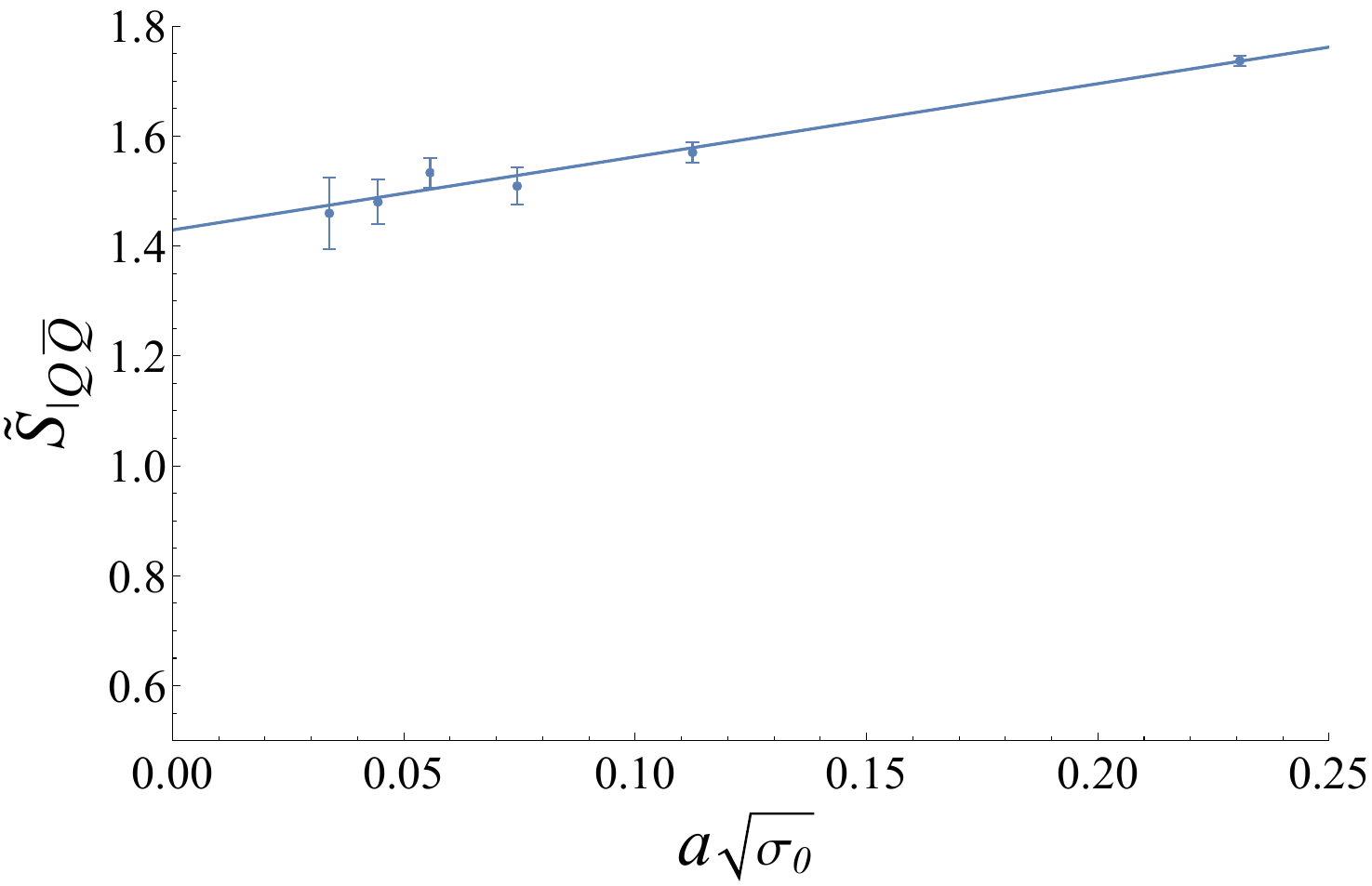}
\includegraphics[width=.49\textwidth]{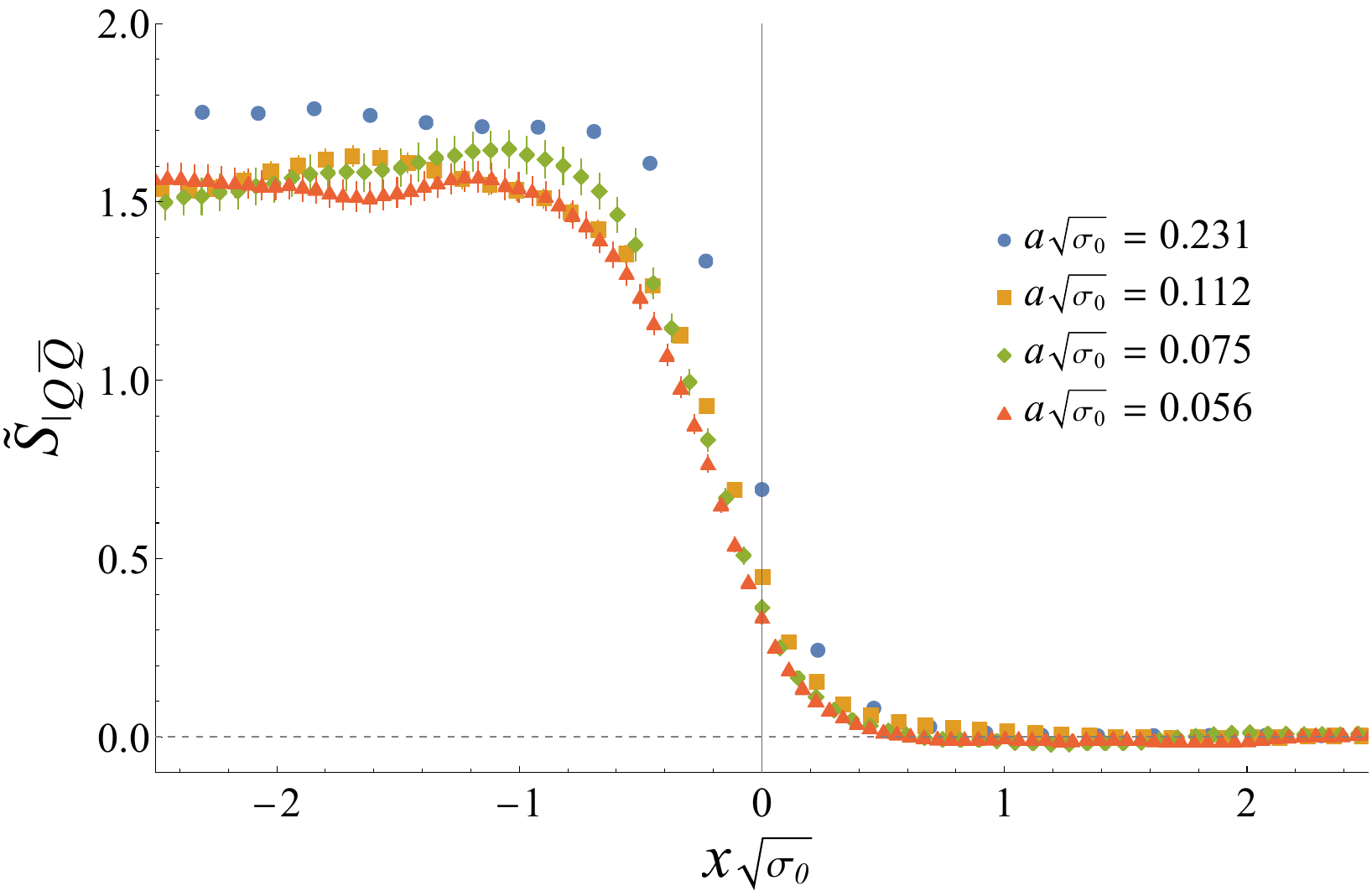}
\\
\caption{\label{fig:su2entropyascaling} 
Left: The Flux Tube Entanglement Entropy (FTE$^2$) $\tilde{S}_{\vert\QQbar}$ in (2+1)D $SU(2)$ Yang-Mills theory as a function of $a$ in the $x\to-\infty$ limit ($x\sqrt{\sigma_0}<-1.8$). Here $x$ is the relative position of the flux tube and region $V$, corresponding to the half-slab geometry shown in Fig. \ref{fig:refinedhalfslab}. Results are shown for a fixed $\QQbar$ distance $L\sqrt{\sigma_0}\approx0.67$, slab width $w\sqrt{\sigma_0}\approx0.22$, and $y=0$.
Right: FTE$^2$ as a function of $x$. Other parameters are the same as in the figure on the left.
}
\end{figure}
\subsection{Results for FTE$^2$}
\label{sec:MCresults}
We performed a scaling study to confirm that FTE$^2$ converges to a finite value in the continuum limit.
We examined  FTE$^2$  at varying values of the lattice spacing $a$ while keeping the quark-antiquark separation and the size of region $V$ fixed in physical ($1/\sqrt{\sigma}$) units.
The results of this study are displayed in Fig.~\ref{fig:su2entropyascaling} (left) for a region $V$ of width $w\sqrt{\sigma_0}\approx 0.22$ and a quark-antiquark separation $L\sqrt{\sigma_0}\approx 0.67$.
As one can see in the figure,  \FTEE is approximately linear in the lattice spacing and has a finite positive value when extrapolated to the continuum limit; the results with lattice spacings $a\sqrt{\sigma_0}\leq 0.1$ agree within the statistical uncertainty.

The values plotted in Fig.~\ref{fig:su2entropyascaling} (left) are obtained by averaging the values at sufficiently large-negative $x\sqrt{\sigma_0}< -1.8$.
The full dependence of \FTEE on the $x$ coordinate is shown in Fig.~\ref{fig:su2entropyascaling} (right), displaying how the \FTEE increases from zero to its asymptotic value $x\to -\infty$ as the flux tube is gradually cross-cut by the half-slab region $V$.
This profile is indicative of the finite width of the flux tube, becoming independent of the scale as  $a\to 0$.
The central values of \FTEE in this region show fluctuations correlated for adjacent points in $x$.
These fluctuations do not exhibit any particular pattern and vary for each ensemble, and are therefore most likely to be of stochastic nature.

\begin{figure}[ht!]
\centering
\includegraphics[width=.49\textwidth]{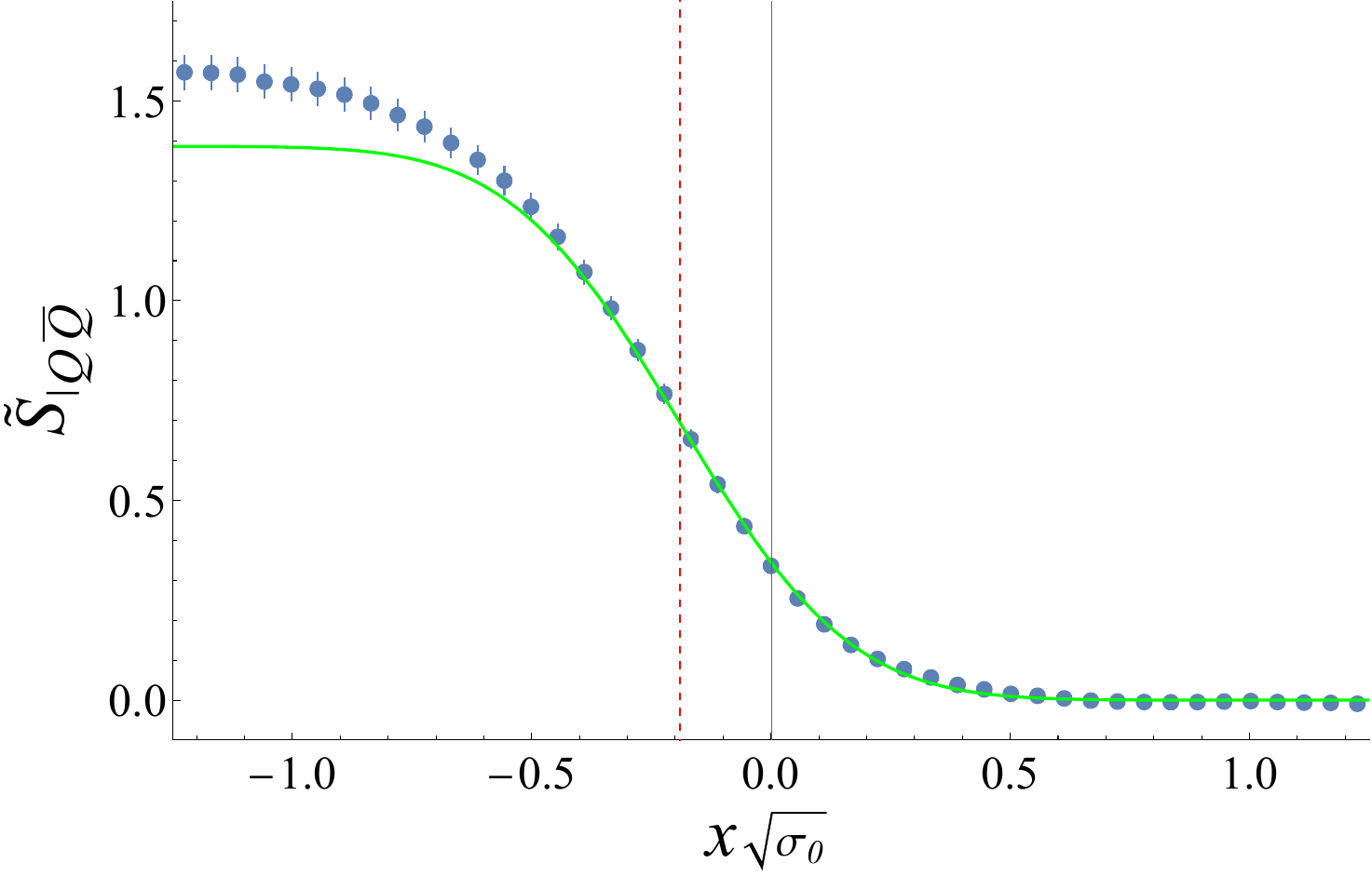}\\
\caption{\label{fig:errorFunctionComp}
\FTEE as a function of $x$ obtained at $a\sqrt{\sigma_0}=0.056$ for $L=12a$,  $w=4a$, and $y=0$.
The green solid line is obtained by fitting the $x\sqrt{\sigma_0}>-0.22$ data to an error function normalized to $2\log N_c$ ($N_c=2$) 
by varying the offset $x_c$ and the half-width $W$.
The optimal error function curve is centered at $x_c\sqrt{\sigma_0}=-0.19$ indicated by the red dashed line and has half-width $W\sqrt{\sigma_0}\approx 0.3$.
}
\end{figure}

In the remainder of this subsection, we will compare the behavior of \FTEE to the predictions of the string model outlined in Section \ref{sec:models}. 
First, we examine the behavior of \FTEE as a function of $x$, which transitions from the completely cross-cut \FTEE value at $x\to -\infty$ to zero at $x\to\infty$.
To study this profile in detail, we focus on the $a\sqrt{\sigma_0}=0.056$ data which are an optimal combination of statistical precision, proximity to the continuum limit, and resolution in $x$. The result is shown in Fig.~\ref{fig:errorFunctionComp}. The value of the completely cross-cut \FTEE ($x\to -\infty$) indicates that it must be dominated by the internal (color) contribution $2\log N_c\approx1.39$ for string length $L\sqrt{\sigma_0}=0.67$.
We also observe that \FTEE at $x\gtrsim0$ closely resembles the error function.
As  anticipated previously from our discussion of Eq.~(\ref{eq:intersectionProbability}), this follows from assuming a Gaussian probability distribution for the transverse deviation of the string and neglecting the vibrational contribution\footnote{
    We do not have a model for the $x$-dependence of the vibrational contribution to FTE$^2$.
    Since it is only a small fraction of \FTEE at $L\sqrt{\sigma_0}=0.67$, we neglect it in analyzing the profile of \FTEE to find its center and width.}.

The $x$ dependence is not symmetric around $x=0$ but somewhat shifted towards negative $x$, which appears contrary to the thin string picture.
Indeed at $x=0$, an infinitely-thin string should have $\approx\frac12$ probability to intersect region $V$, therefore the internal entropy in Eq.~(\ref{eq:intersectionProbability}) should contribute half of the asymptotic value.
However, the \FTEE at $x=0$ is substantially less than that, reaching 50\% only at $x\sqrt{\sigma_0}\approx-0.19$. 
To understand this better, we fit \FTEE data for $x\sqrt{\sigma_0}>-0.22$ to an error function $\sim (2\log N_c) \erf\big((x_c-x)/(W\sqrt{2})\big)$ and find its center at $x_c\sqrt{\sigma_0}=-0.19$ and its Gaussian half-width $W\sqrt{\sigma_0}\approx0.3$.
The half-width $W$ of the profile matches the expectation for the profile of the flux tube~\cite{Caselle:2012rp,Caselle:2021eir}, 
while the displacement $x_c$ of the profile center might be associated with the intrinsic width of the string, which is neglected in the thin string approximation.

\begin{figure}[ht!]
\centering
\includegraphics[width=.49\textwidth]{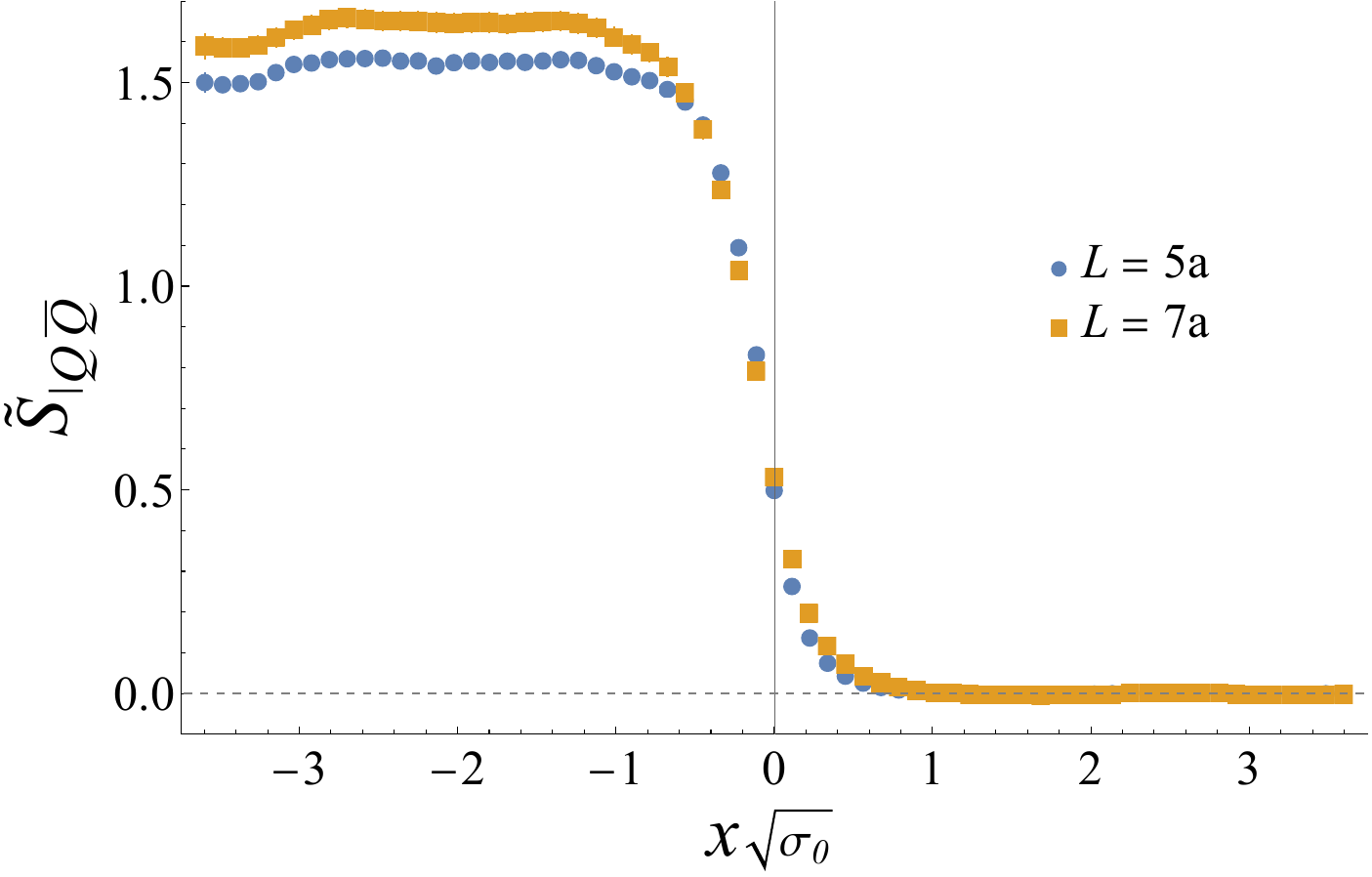}
\includegraphics[width=.49\textwidth]{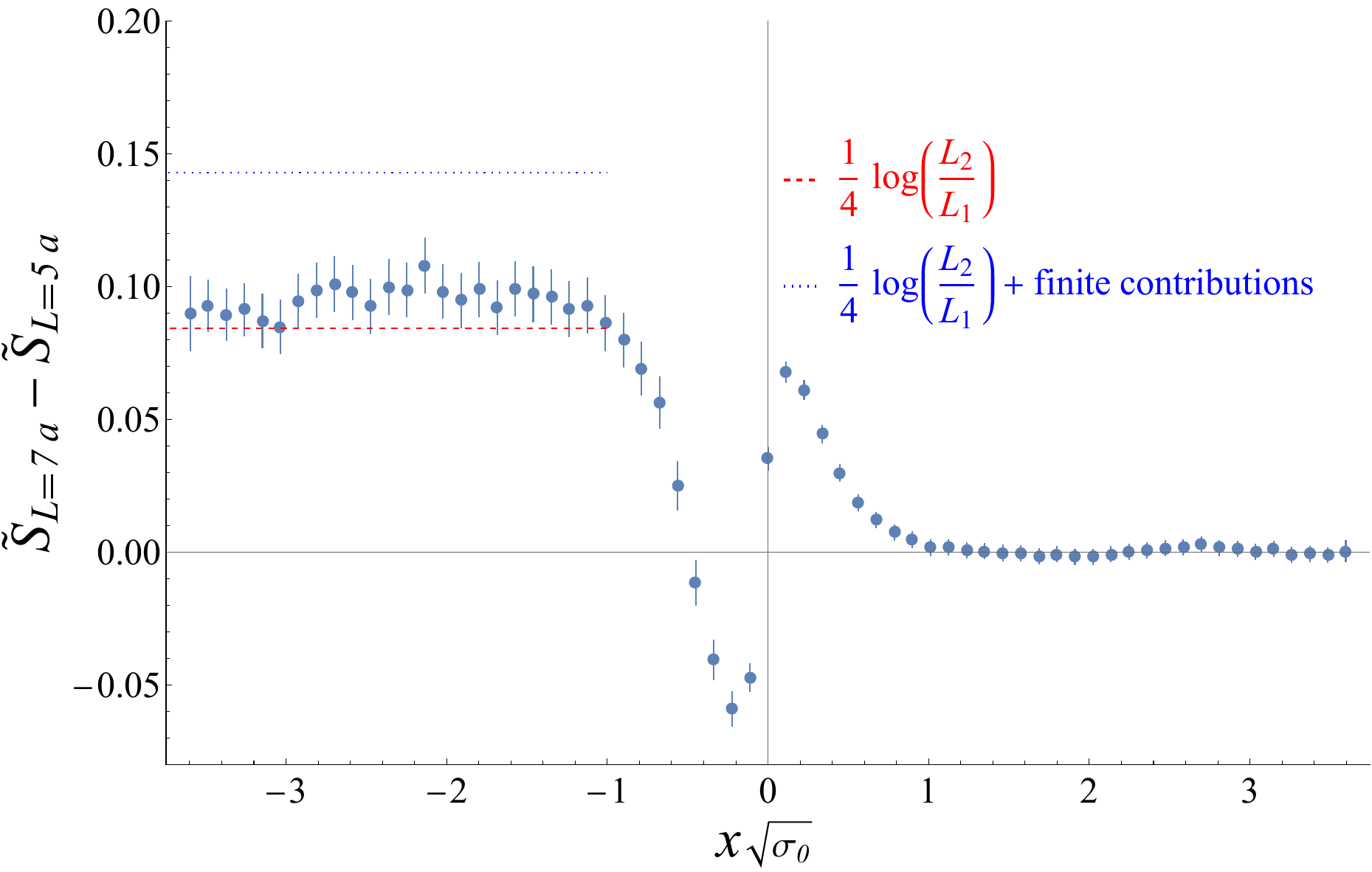}\\
\caption{\label{fig:entropystringlengthdependence}
Left: FTE$^2$ as a function of $x$ for different values of flux tube length $L$. 
Results are shown for $a\sqrt{\sigma_0}=0.1124$, $w=3a$, and $y=0$. Points are offset horizontally for clarity. Right: The difference $\big[\tilde{S}_{\vert\QQbar}(L=7a) - \tilde{S}_{\vert\QQbar}(L=5a)\big]$ of FTE$^2$ shown in the left plot. 
String model vibrational entropy calculations given in Section \ref{sec:models} for completely cross-cut strings ($x\to-\infty$), are shown with dashed (length scaling term) and dotted (full string model) lines.}
\end{figure}

Next, we examine the \FTEE dependence on the length of the flux tube $L$.
Our data at $a\sqrt{\sigma_0}\approx0.112$ indicates that the completely cross-cut \FTEE value increases by $\sim7\%$ 
as the length of the flux tube is increased by $40\%$, as seen in Fig.~\ref{fig:entropystringlengthdependence} 
 (left). In Fig.~\ref{fig:entropystringlengthdependence} (right), we see the difference of \FTEE computed with lengths $L=5a$ and $L=7a$ and fixed $w=3a$ as a function of $x$.
We expect the internal (color) entropy to cancel in this difference for $x\to\infty$ and mostly cancel for intermediate $x$.
As $x\to -\infty$, this difference agrees very well with the analytic prediction for the divergent logarithmic growth of Renyi vibrational entropy $S^{(2)}\sim\frac{1}{4} \log(L/\epsilon)$ of a thin string shown as a horizontal line. However, the model also predicts that the finite part must depend on the fraction of the string $w/L$ in region $V$-shown as a horizontal line in Fig.\ref{fig:entropystringlengthdependence} (right)--which clearly does not agree with the lattice result.

Further examining the $x$-dependence of the difference in Fig.\ref{fig:entropystringlengthdependence} (right) reveals non-monotonic behavior: a trough and a peak in the $|x\sqrt{\sigma_0}| < 1$ interval.
This interval is comparable to the width of the flux tube, and the non-monotonic behavior has a natural qualitative explanation in terms of the internal entropy. 
Indeed, since the prediction for the internal entropy in Eq.~(\ref{eq:intersectionProbability}) depends on the probability of the string crossing into region $V$, it must change as the width of the flux tube increases with  length~\cite{Luscher:1980iy}.
In other words, the wider profile of a longer flux tube allows the thin string to deflect into region $V$ more often when $x$ is positive, and conversely, avoid region $V$ (see Fig.~\ref{fig:refinedhalfslab}) more often when $x$ is negative, compared to a shorter and narrower flux tube.
This is in good agreement with the behavior seen in Fig.\ref{fig:entropystringlengthdependence} (right).

\begin{figure}[ht!]
\centering
\includegraphics[width=.49\textwidth]{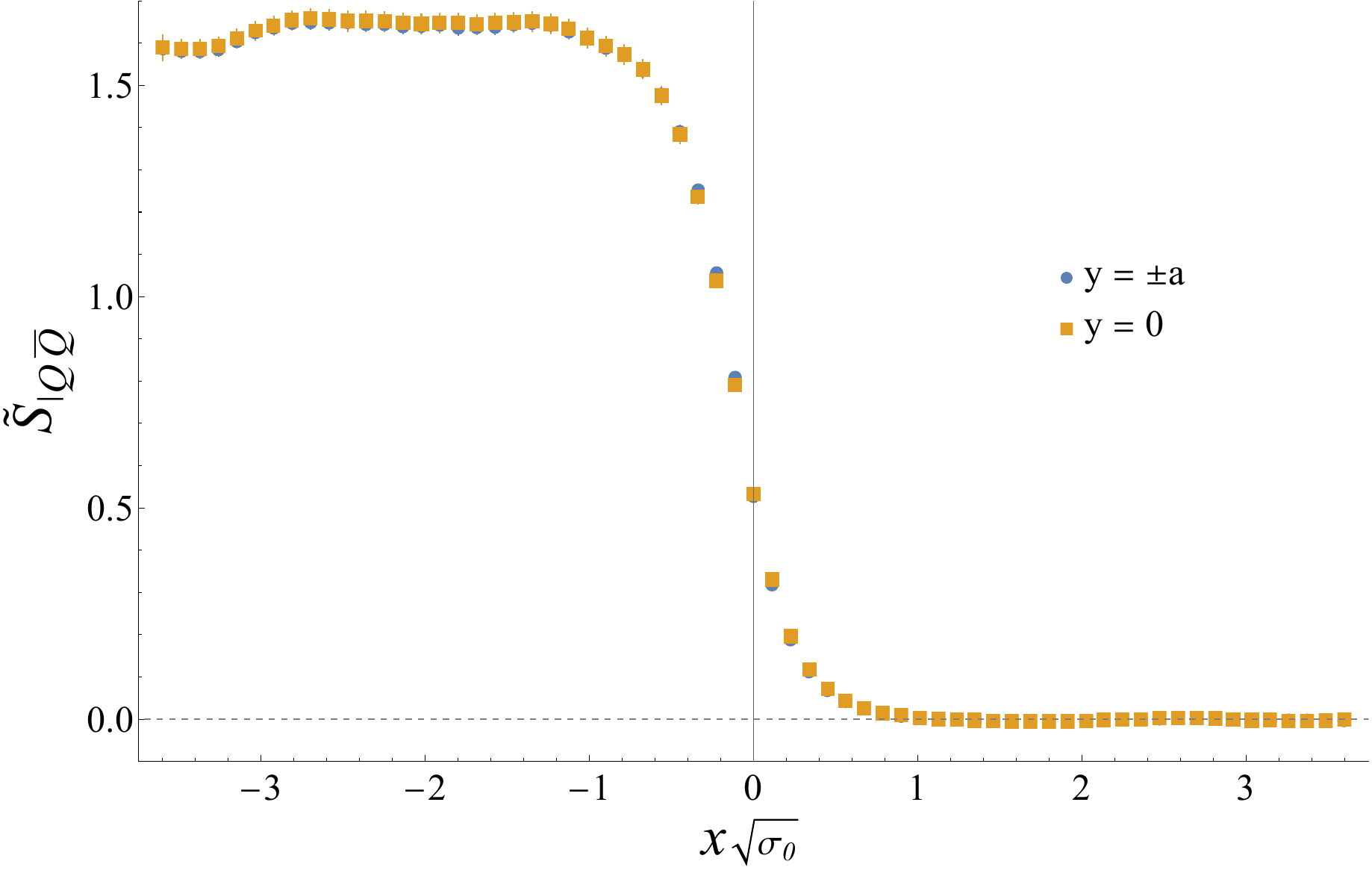}
\includegraphics[width=.49\textwidth]{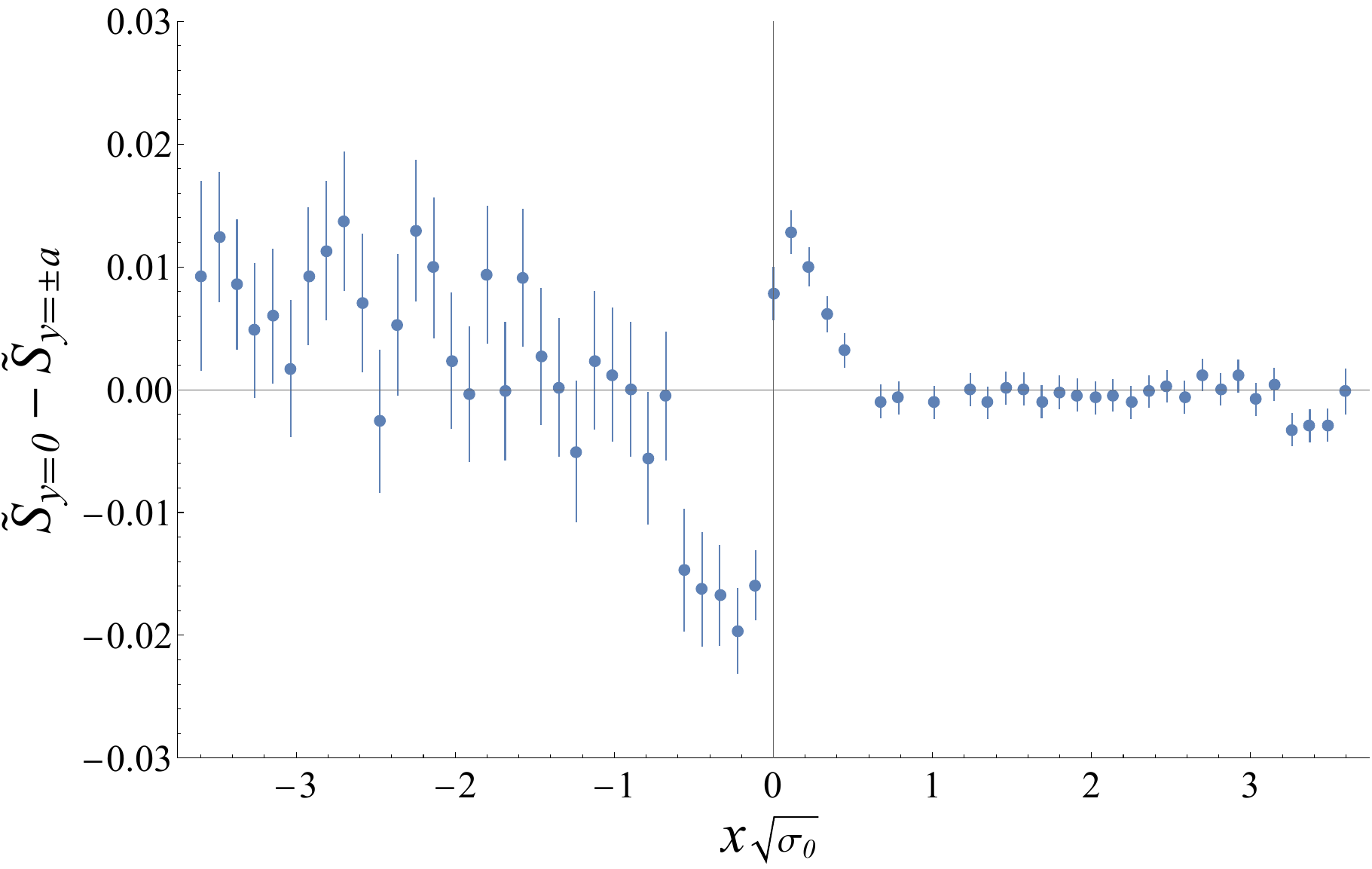}\\
\caption{\label{fig:su2entropylocationdependence}
Left: Dependence of FTE$^2$ on the position $y$ relative to the center of region $V$. 
Results are shown for $a\sqrt{\sigma_0}=0.1124$.
The $\QQbar$ distance is $L = 7a$. Points are offset slightly horizontally for clarity.
Right: Difference in FTE$^2$ between a string of length $L=7a$ straddling region $V$ of width $w=3a$ at $y=0$ (centered) and $y=\pm a$ (off-center).}
\end{figure}

The difference between the thin string vibrational entropy and the lattice data noted above can be further examined by changing either the width $w$ or the location $y$ of region $V$ relative to the center of the string and keeping all else equal.
The difference due to the logarithmic growth of the Renyi entanglement entropy $S^{(2)} \sim\frac{1}{4}\log(L/\epsilon)$ is eliminated if $L$ is kept constant, and any variation will therefore be solely due to the $w$- or $y$-dependence of the finite correction.
This correction depends only on the geometric proportions $(w/L)$ and $(y/L)$ of the $V/\bar V$ partitioning of the flux tube, and we expect the behavior to be similar to the calculations in Sec.~\ref{sec:models} shown in Fig.~\ref{fig:wLocAnalyticScaling}.

The comparison of \FTEE for $y=0$ and $y\ne0$ is shown in Fig.~\ref{fig:su2entropylocationdependence}, which are very close, and their difference is consistent with zero within the uncertainties.
This finding does not agree with the expectations of the string model outlined in Sec.~\ref{sec:models}.
This disagreement is not particularly surprising because, as noted in Sec.~\ref{sec:models}, the finite correction term is extracted in the limit $L/\epsilon\to\infty$, while the length of the flux tube in our lattice computations is very short, as evidenced by how small the vibrational contribution to the \FTEE is.
For intermediate $x$ (partially cross-cut flux tube), the difference of \FTEE for $y=0$ and $y=\pm a$ in Fig.~\ref{fig:su2entropylocationdependence} (right) exhibits non-monotonic behavior similar to Fig.~\ref{fig:entropystringlengthdependence}.
This too can be qualitatively understood as the effect of the flux tube width, which is maximal in the middle ($y=0$) and decreases towards its ends ($y\ne0$).
Therefore, the midpoint of the vibrating string ($y=0$) is more likely to cross into region $V$ if $x$ is positive and avoid it if $x$ is negative, compared to the points of the string closer to its fixed ends ($y\ne0$).

\begin{figure}[ht!]
\centering
\includegraphics[width=.49\textwidth]{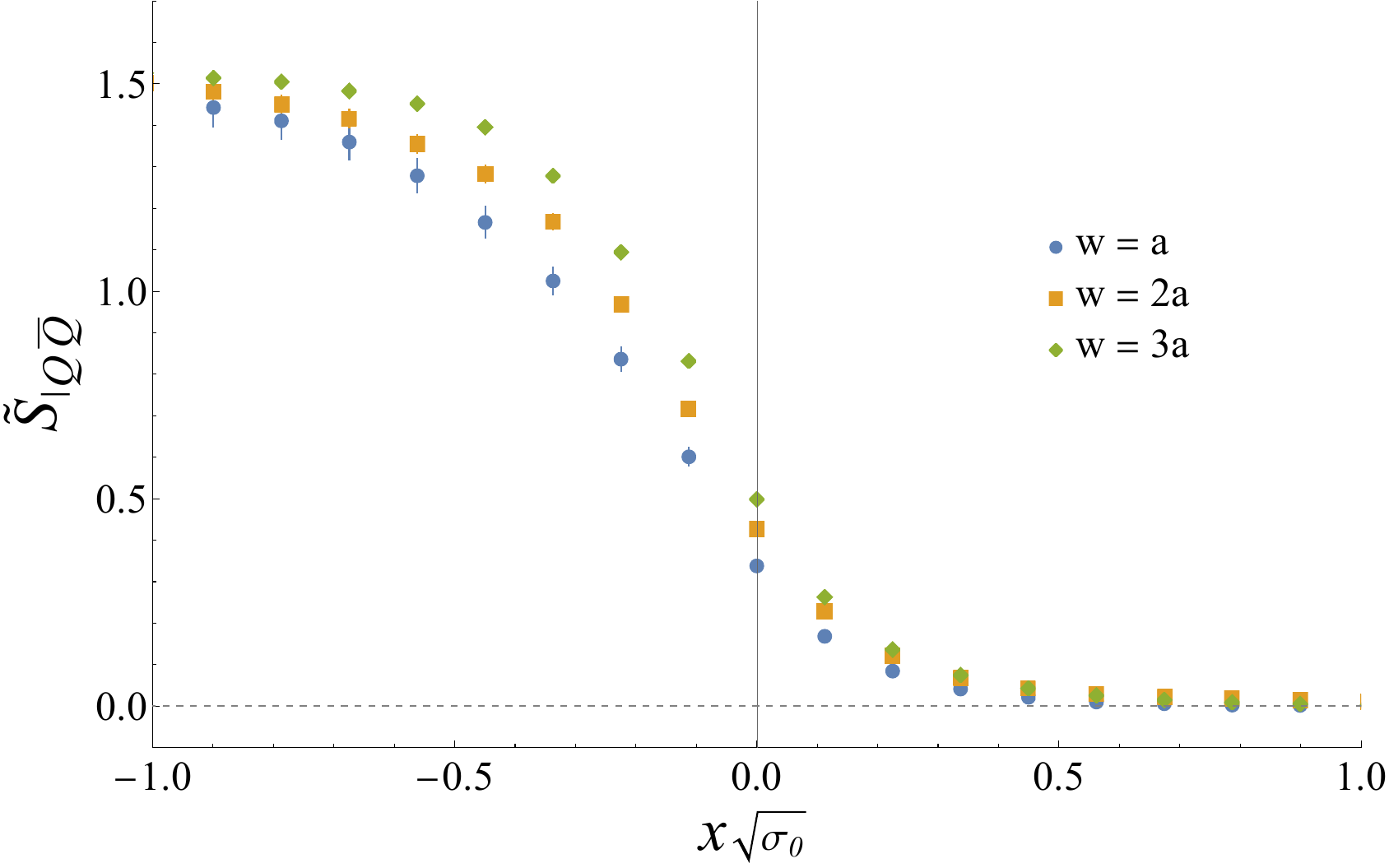}\\
\caption{\label{fig:su2entropyregionwidthdependence}
The dependence of FTE$^2$ on the width $w$ of region $V$ for $a\sqrt{\sigma_0}
\approx0.112$ and the flux tube length $L = 5a$.
$w=a$ and $w=3a$ ensembles were calculated with $y=0$, while $w=2a$ was calculated at $y=\pm a/2$. Since the length dependence is minimal, we can neglect the difference between $y=0$ and $y=\pm a/2$.}
\end{figure}
Studying the $w$-dependence of \FTEE is more difficult because for different $w$ it has to be evaluated on uncorrelated Monte Carlo ensembles, and one cannot effectively eliminate stochastic noise as in the cases of the $L$- and $y$-dependence.
Also, the granularity of the lattice limits the choice of $w$ values with matching $y$ values, for example, with $y=0$ that corresponds to the center of the flux tube.
However, since the $y$-dependence is very weak, as we saw in Fig.~\ref{fig:su2entropylocationdependence} (right), we can neglect the difference between $y=0$ and $y=\pm\frac a2$, and compare \FTEE for $w=(1,2,3)a$ in Fig.~\ref{fig:su2entropyregionwidthdependence}. 
For a completely cross-cut flux tube ($x\to-\infty$), only the finite contribution that depends on $(w/L)$ should contribute.
Our present statistics is not sufficient to resolve any difference there.
However, there is a distinct difference in the behavior of \FTEE on $w$ at intermediate $x$, where the flux tube is partially cut.
At $x=0$, we find that \FTEE exhibits roughly linear growth with $w$, which is most pronounced near $x\sqrt{\sigma_0}\sim -0.2$.
This behavior is reminiscent of the area law for the entanglement entropy, with the assumption that the only relevant part of the boundary between $V$ and $\bar V$ is where it spatially overlaps the flux tube.

Overall, our results indicate that while key features of our lattice data for \FTEE in (2+1)D Yang-Mills theory are captured by the sum of the internal entropy, whose structure is anticipated by (1+1)D analytical results, and the vibrational entropy estimated from the thin string model, a quantitative comparison will require refinements to our models of both. 
These can be constrained by extending our results to $N_c > 2$, to (3+1)D, and perhaps by simulations at $T > T_c$. 

\section{Summary and outlook
  \label{sec:summary_outlook}}
In this paper we introduced a novel Flux Tube Entanglement Entropy (FTE$^2$), a gauge-invariant and UV finite physical measure of the entanglement entropy of regions within and surrounding the color flux tube formed by a static quark-antiquark pair in pure Yang-Mills theory.
In the first part of this work, we discussed at length the desired properties of FTE$^2$ within the larger context of the gauge invariance of the entanglement entropy, focusing in particular on the dependence of the entanglement entropy on the choice of the center of the algebra employed in such a computation. We outlined the arguments in the literature discussing both the general algebraic construction of the entanglement entropy and the extended lattice construction that is equivalent to the choice of the electric-center choice.  We concluded that since FTE$^2$ is manifestly gauge-invariant, it should also be center-independent. 

We then laid out our procedure for extracting  FTE$^2$ on the lattice using correlators of Polyakov loops and the replica method. Our framework allows us to study FTE$^2$ for a slab of arbitrary width for the cases where the slab fully or partially cross-cuts the flux tube, or does not cross it at all. 
We postulated a bifurcation of the entanglement entropy into the sum of the entanglement due to colorful degrees of freedom inside the flux tube and entanglement due to the transverse vibrations of the color flux tube.
We performed explicit computations of FTE$^2$ in (1+1)D Yang Mills theory to model the internal entropy. The results of this computation are interesting in their own right and will be reported separately. We also employed the simple thin string model to model the vibrational entropy. In the long string limit, we showed that the vibrational entropy of the flux tube exhibits the same logarithmic length dependence expected from 2D conformal field theory.

We then discussed the results of Monte Carlo simulations of FTE$^2$ on the lattice for (2+1)D $SU(2)$ pure gauge theory. We first confirmed that FTE$^2$ is UV finite in the continuum limit. Examining FTE$^2$ when the string is fully cut, we found that its value lies above our estimate for the internal entropy, indicating contributions from both the internal and vibrational entropy. We further found that FTE$^2$, as a function of the overlap of the slab with the flux tube, is qualitatively well-described by an error function shape expected from the Gaussian profile of small deflections of the flux tube when the flux tube is near the boundary. In particular, we find that the length dependence term of the vibrational entropy accurately describes the length dependence we see on the lattice. We expect that the string portion term of the vibrational entropy does not apply to flux tubes of the length explored in this paper, and and indeed found that its predictions do not match the Monte Carlo data.
Further studying FTE$^2$, we found that it is well described by the predictions of our internal entropy model when the flux tube is partially cut, with a slight caveat; FTE$^2$ behaves as if it were a thin string further away from the boundary than it is, by a distance on the order of around $1/3$ the width of the flux tube.

This work leaves us with many further directions to explore.
One of the questions not fully resolved by this work is the accuracy of the postulated bifurcation of the entanglement entropy and the applicability of the thin string model.
There are many ways to model Flux Tube Entanglement Entropy more rigorously.
One could also, as we plan to do, vary the number of colors, testing Eq.~(\ref{eq:intersectionProbability}).
Another way to test Eq.~(\ref{eq:intersectionProbability}) is to vary the number of flux tube boundary crossings. 
In (1+1)D, there are no transverse dimensions and the flux tube has no intrinsic width.
It would be reasonable to suspect non-trivial behavior of FTE$^2$ as the internal structure of the flux tube takes on another dimension in (2+1)D.
One could place multiple slabs in between quark and antiquark sources in (2+1)D to further test our understanding of the flux tube's entropic dynamics.

Changing the number of dimensions of space, as well as performing a finite temperature study to explore the behavior of FTE$^2$ above the deconfinement transition temperature in pure Yang-Mills represent obvious directions for future work.
A further interesting possibility is to explore how FTE$^2$ behaves when one changes the  number of static quark sources. This is especially timely given the interest in empirical understanding the role of baryon junctions in carrying baryon number~\cite{STAR:2024lvy,ALICE:2013yba,Lewis:2022arg}. Their entropic dynamics might provide useful input into constructing useful effective models of baryons that can potentially help distinguish between different mechanisms proposed in the literature~\cite{Rossi:1977cy,Kopeliovich:1998ps,Garcia-Montero:2024jev,Frenklakh:2024mgu,Komargodski:2024swh}.

While there are many potentially fruitful directions to take with Monte Carlo simulations, there is also more work needed to be done in applying effective string models that capture the dynamics of FTE$^2$.
For this work, we used a simple thin string model with no corrections and it will be important to develop this study in the context of the long string EFT. 
One of the issues that will likely persist regardless of model is the offset of FTE$^2$ from predictions by a distance on the order of around $2/3$ the  width of the flux tube.
Our results seem to suggest that the flux tube contributes to the  entanglement entropy of a region only when its entire intrinsic width  lies in that region, raising significant questions about the internal entropic structure of color flux tubes and where their degrees of freedom lie. All of these issues will be addressed in forthcoming work.

\section*{Acknowledgements}
R.A is supported by the Simons Foundation under Award number 994318 (Simons Collaboration on Confinement and QCD
Strings). 
S.S. is supported by NSF supported by the National Science Foundation under Award PHY-2412963. 
In addition, R.A. is supported in part by the Office of Science, Office of Nuclear Physics,
U.S. Department of Energy under Contract No. DEFG88ER41450 and by the National Science Foundation under Award
PHY-2412963.
R.V is supported by the U.S. Department of Energy, Office of Science under contract DE-SC0012704. 
R.V's work on quantum information science is supported by the U.S. Department of Energy, Office of Science, National
Quantum Information Science Research Centers, Co-design Center for Quantum Advantage (C$^2$QA) under contract number
DE-SC0012704. 
R.V. was also supported at Stony Brook by the Simons Foundation as a co-PI under Award number 994318 (Simons Collaboration on Confinement and QCD Strings). 
The authors thank Stony Brook Research Computing and Cyberinfrastructure and the Institute for Advanced Computational
Science at Stony Brook University for access to the Seawulf HPC system, which was made possible by grants from the
National Science Foundation (awards 1531492 and 2215987) and matching funds from the Empire State Development’s Division
of Science, Technology and Innovation (NYSTAR) program (contract C210148).

\appendix

\section{Vibrational entanglement entropy of a color flux tube
  \label{app:thin_string}}
In this Appendix, we will discuss how to model the vibrational entanglement entropy of a color flux tube contributing to
FTE$^2$. 
For the vibrational entropy, we employ a Gaussian approximation of the relativistic string valid for small fluctuations
$x_\perp$ of the relativistic string model~\cite{Luscher:1980iy}. 
Our model does not fully capture the string theory formalism since our string cannot ``turn back''.
The string's transverse displacement is instead a function of the longitudinal coordinate along the string, and we can
parameterize the string in two spatial dimensions as \cite{Luscher:1980iy}, 
\begin{equation}
    \vec{x}(s)=\Big(x_\perp(s),\frac{L}{\pi}s\Big)\,, 0\leq s \leq\pi\,.
\end{equation}
The ``thin string" Hamiltonian is
\begin{equation}
    H=M^2L+\frac{\pi}{2M^2L}\int\limits_{0}^{\pi} ds (p^2+M^4x'(s)^2)\,,
    \label{eqn:Hamiltonian_thinstring2}
\end{equation}
where $p$ and $x'$ are understood to be measured in the direction perpendicular to the string, $x' \equiv
\frac{dx}{ds}$, and $p \equiv -i\frac{\delta}{\delta x(s)}$, and $M^2$ represents the string tension, which is inversely
proportional to the width of the string.

To compute the vibrational entanglement entropy, we first express the string as a discrete set of points $x(s_0)$
through $x(s_N)$, each with transverse displacement $x_{\perp}(s_i)$.
We will work with one transverse dimension\footnote{As the longitudinal coordinate $x_\parallel $ is proportional to
$s$, when we write $x(s)$ throughout this section, we will always be referring to $x_\perp(s)$.}.
The string is assumed to interpolate linearly between the basis points.
To find $p$, we expand 
\begin{align}
\label{eqn:oscillatormodes}
x(s) &= \sum\limits_{i} a_i x_i(s)\,,\\
\rm{with}\,\,\int ds\, [x(s)]^2 &= \sum\limits_{i,j} a_ia_j g_{ij}\,,
\end{align}
where $g_{ij} \equiv \int ds\,x_i(s)x_j(s)$ is the ``metric tensor'' for the basis $\{x_i\}$. 
In principle, this basis can be completely arbitrary; for our purposes, we will use 
\begin{equation}
    x_i(s)= \begin{Bmatrix} 1-\frac{N}{\pi}\abs{s-\pi(\frac{i}{N})} \text{ if, } \pi(\frac{i-1}{N})<s<\pi(\frac{i+1}{N}) \\ \\
    0,\,\, \text{ otherwise} \end{Bmatrix}
\end{equation}
with $1\leq i\leq N-1$ and endpoints $x_0(s)=x_N(s)=0$.
For this basis \footnote{The diagonal $g_{ij}=\epsilon\delta_{ij}$ produces results that agree with this definition in the continuum limit.},
\begin{align}
    g_{ij} = \begin{Bmatrix} \frac{2\pi\epsilon}{3N} &\text{ if } i=j \\ \frac{\pi\epsilon}{6N} &\text{ if } i=j\pm1 \\ 0 &\text{ otherwise} \end{Bmatrix} .
\end{align}
The conjugate $p(s) = -i \delta/\delta x(s)$ is a ``co-vector'' $[x(s), p(s')] = i\delta(s,s')$, and therefore its expansion in the co-vector components $\delta/\delta a_i$ is
\begin{equation}
\label{eq:gmnmodes}
\int ds\, [p(s)]^2 = -\sum\limits_{i,j} \frac{\delta}{\delta a_i}\frac{\delta}{\delta a_j} 
g_{ij}^{-1}\,.
\end{equation}
We now have the necessary ingredients to diagonalize the Hamiltonian. We first write 
\begin{equation}
   H= \vec{a}^T X\,\vec{a}+\vec{p}^T Q\,\vec{p} \,,
   \label{Hammodes}
\end{equation}
putting in quadratic form both the ``position" coordinates and their conjugate momenta, with  $\vec{a}=(a_1\dots a_i\dots a_{N-1})$  and the vector $\vec{p}=-i(\frac{\delta}{\delta a_1}\dots \frac{\delta}{\delta a_i}\dots  \frac{\delta}{\delta a_{N-1}})$. 
The matrix $Q$ has components $Q_{ij} = \frac{\pi}{2M^2L} g^{-1}_{ij}$, as seen from Eqs.~(\ref{eq:gmnmodes}) and ~(\ref{eqn:Hamiltonian_thinstring2}). The matrix $X$ can be expressed as
\begin{align}
    X_{ij} = \begin{Bmatrix} \frac{\pi M^2}{L\epsilon} &\text{ if } i=j \\
    \frac{-\pi M^2}{2L\epsilon} &\text{ if } i=j\pm1 \\
    0 &\text{ otherwise} \end{Bmatrix} .
\end{align}
We then decompose $X$ as
\begin{equation}
    X = Y^TY
\end{equation}
and define
\begin{equation}
    b \equiv Ya .
\end{equation}
$p$ is no longer the conjugate momentum to $b$, and we would like to express the Hamiltonian in terms of coordinates and their conjugate momenta. Therefore, defining 
\begin{equation}
    r \equiv (Y^{-1})^T p ,
\end{equation}
the conjugate momentum to $b$, we arrive at
\begin{equation}
    H = \vec{b}^T \vec{b}+\vec{r}^T\, Y Q Y^T \,\vec{r}.
\end{equation}
To diagonalize the non-trivial quadratic term, we define an orthogonal matrix $Z$ such that
\begin{equation}
    D \equiv Z Y Q Y^T Z^T\,,
\end{equation} 
is diagonal. Transforming the momentum and coordinate vectors, respectively, once again as
\begin{equation}
    s \equiv Z r \,,\,\,\,
    c \equiv (Z^{-1})^T b \,,
\end{equation}
we can finally express the Hamiltonian in the quadratic form
\begin{equation}
    H = c^T c + s^T D \,s .
\end{equation}
This Hamiltonian has the ground state 
\begin{equation}
    \Psi(\vec{c})\propto \text{exp}[-\frac{1}{2}c^T D^{-1/2} c] \,,
\end{equation}
which can now be rexpressed in terms of  the original ``position" coordinates $a$ as 
\begin{equation}
    \Psi(\vec{a}) \propto \text{exp}[-\frac{1}{2}a^T {\cal M}\, a] \,,
    \label{posrepfin}
\end{equation}
where 
\begin{equation}
\label{eqn:Mdef}
    {\cal M} \equiv Y^T (YQY^T)^{-1/2} Y \,.
\end{equation}
This matrix  can be broken up into the components
\begin{equation}
    \cal M=\begin{pmatrix} \gamma_{VV} & \gamma_{V\bar{V}} \\ \gamma_{\bar{V}V} & \gamma_{\bar{V}\bar{V}} \,,
    \end{pmatrix}
\end{equation} 
which equivalent to Eq.~9 in \cite{Callan:1994py}. As this form of the matrix suggests, if modes $\vec{a}$ are broken up into modes with support in $V$ and modes with support in $\bar{V}$ in the form of $\begin{pmatrix} a_{V} \\ a_{\bar{V}} \end{pmatrix}$, we can reexpress Eq. (\ref{posrepfin}) as
\begin{equation}
    \Psi(\vec{a}) \propto \text{exp}\Big[-\frac{1}{2}(a_{V} a_{\bar{V}})\cal  M \begin{pmatrix} a_{V} \\ a_{\bar{V}} \end{pmatrix}\Big] \,,
\end{equation}
or equivalently, 
\begin{equation}
    \Psi_0 \propto \text{exp} \Big[ -\frac{1}{2}(a_{V}a_{\bar{V}}) \begin{pmatrix} \gamma_{VV} & \gamma_{V\bar{V}} \\ \gamma_{\bar{V}V} & \gamma_{\bar{V}\bar{V}} \end{pmatrix}\begin{pmatrix} a_{V} \\ a_{\bar{V}} \end{pmatrix}\Big]\,.
\end{equation} 
Without loss of generality, we can set $\gamma_{V\bar{V}}=\gamma_{\bar{V}V}^T$. 
Following \cite{Callan:1994py}, we can now perform Gaussian integration over the complement region 
$\bar{\cal A}$. We first write
\begin{equation}
    \hat{\rho}(a_1,a_2) = \int \mathcal{D} a_{\bar{V}} \psi_0(a_{1}a_{\bar{V}})\psi_0(a_{2}a_{\bar{V}}) ,
\end{equation}
with $a_1$ and $a_2$ in $V$. Completing the square, and performing the Gaussian integral, we obtain
\begin{equation}
    \hat{\rho}(a_1,a_2) = -\int \mathcal{D} a_{\bar{V}} \exp(C^T \gamma_{\bar{V}\bar{V}}\,C-a_1^T(A)\,a_1-a_2^T(A)\,a_2-a_1^T(4B)\,a_2)
\end{equation}
where 
\begin{eqnarray}
    A &=& 2\,(\gamma_{VV}-\frac{1}{2}\gamma_{V\bar{V}}(\gamma_{\bar{V}\bar{V}})^{-1}\gamma_{\bar{V}V})\nonumber\,,\\ 
    B &=& -\frac{1}{2}\gamma_{V\bar{V}}(\gamma_{\bar{V}\bar{V}})^{-1}\gamma_{\bar{V}V}\nonumber\,,\\ 
    C &=& a_{\bar{V}}+\frac{1}{\sqrt{2}} (\gamma_{\bar{V}\bar{V}}^{-1})^T\gamma_{\bar{V}V}a_{1}+\frac{1}{\sqrt{2}} (\gamma_{\bar{V}\bar{V}}^{-1})^T\gamma_{\bar{V}V}a_{2} \,.
\end{eqnarray}
This reduced density matrix can be reexpressed as 
\begin{equation}
    \hat{\rho}(a_1,a_2)\propto \text{exp} \Big[ -\frac{1}{2}(a_{1}a_{2}) \begin{pmatrix} A & 2B \\ 2B & A \end{pmatrix}\begin{pmatrix} a_{1} \\ a_{2} \end{pmatrix} \Big] .
    \label{densmat}
\end{equation}
We can then express $\rho^2$ as the functional integral
\begin{equation}
    \text{tr}\rho^2=\int \mathcal{D}\phi\, \text{exp}\Big[ -(a^1 a^2) \mathcal{M}_2 \begin{pmatrix} a^{1} \\ a^{2} \end{pmatrix}\Big]\,,
    \label{rhosquared}
\end{equation}
where $\mathcal{M}_2$ is the supermatrix
\begin{equation}
    \mathcal{M}_2=\begin{pmatrix} A & 2B \\ 2B & A \end{pmatrix}.
\end{equation}
Note that $\mathcal{M}_2$ differs from the general formula for $\mathcal{M}_n$ used in \cite{Callan:1994py}.
Following the prescription in \cite{Callan:1994py}, we use the expression 
\begin{equation} 
\text{tr}\rho^n=1/\sqrt{\text{det}\mathcal{M}_n}\,,
\label{tracesuper}
\end{equation} 
to then compute the Rényi entropy for various numbers of modes $N$, normalizing so $\text{tr}\rho=1$. One can also compute the von Neumann entanglement entropy at this point, defining 
\begin{equation}
    M'\equiv \gamma_{V\bar{V}}(\gamma_{\bar{V}\bar{V}})^{-1}\gamma_{\bar{V}V}
\end{equation} and 
\begin{equation}
    N'\equiv \gamma_{VV}+\gamma_{V\bar{V}}(\gamma_{\bar{V}\bar{V}})^{-1}\gamma_{\bar{V}V}.
\end{equation}
Following \cite{Bombelli:1986rw}, we can write the Von-Neumann entanglement entropy as 
\begin{equation}
    S= -\sum\limits_{n} \frac{\mu_n\text{log}\mu_n+(1-\mu_n)\text{log}(1-\mu_n)}{1-\mu_n}
\end{equation}
where $\mu_n$ represent the unique positive solution of $\lambda_n=4\mu_n/(1-\mu_n)^2$, and $\lambda_n$ represent the eigenvalues of $\Lambda\equiv M'^{-1}N'$.
As discussed in the main text, the thin string model produces the following apparent behavior of the Rényi and von Neumann entropies for the fully cut string:
\begin{equation}
    S^{(2)}=\frac{1}{4}\text{log}(N)+\text{finite corrections}\,,
\end{equation}
\begin{equation}
    S=\frac{1}{3}\text{log}(N)+\text{finite corrections}\,.
\end{equation}
In the main text, $N$ is replaced with $L/\epsilon$, where $L$ is the length of the string and $\epsilon$ is the distance between consecutive basis points.
These two formulations are equivalent.

\bibliography{bib}
\end{document}